\documentclass[twocolumn]{revtex4-1}       

\usepackage{graphicx}                
\usepackage{amssymb}
\usepackage{amsmath}
\usepackage{color}

\newcommand*{\TRANS}{^{\mkern-1.5mu\mathsf{T}}}
\newcommand*{\VEC}[1]{\boldsymbol{#1}}
\newcommand*{\OP}[1]{{\cal{#1}}}

\newcommand*{\EPS}{\varepsilon}

\DeclareMathOperator\arccosh{arccosh}

\begin{document}

\title{Entropy production of a Brownian ellipsoid in the overdamped limit}

\author{Raffaele Marino}
\author{Ralf Eichhorn}
\email{eichhorn@nordita.org}
\affiliation{
Nordita \\
Royal Institute of Technology and Stockholm University \\
Roslagstullsbacken 23, SE-106 91 Stockholm, Sweden
}
\author{Erik Aurell}
\email{eaurell@kth.se}
\affiliation{
Dept. of Computational Biology and ACCESS Linnaeus Centre and Center for Quantum Materials,
KTH -- Royal Institute of Technology,  AlbaNova University Center, SE-106 91~Stockholm, Sweden\\
Depts. Information and Computer Science and Applied Physics, Aalto University, Espoo, Finland
}

\begin{abstract}
We analyze the translational and rotational motion of an ellipsoidal Brownian particle
from the viewpoint of stochastic thermodynamics.
The particle's Brownian motion is driven by external forces and torques and takes place in
an heterogeneous thermal environment where friction coefficients and (local) temperature
depend on space and time.
Our analysis of the particle's stochastic thermodynamics
is based on the entropy production associated with single particle trajectories.
It is motivated by the recent discovery that the overdamped limit of vanishing
inertia effects (as compared to viscous fricion) produces a so-called ``anomalous''
contribution to the entropy production, which has no counterpart in the overdamped
approximation, when inertia effects are simply discarded.
Here, we show that rotational Brownian motion in the overdamped limit
generates an additional contribution to
the ``anomalous'' entropy. We calculate its specific form by performing a
systematic singular perturbation analysis for the generating function of the
entropy production. As a side result, we also obtain the (well-known) equations
of motion in the overdamped limit. We furthermore investigate the effects of
particle shape and give explicit expressions of the ``anomalous entropy''
for prolate and oblate spheroids and for near-spherical Brownian particles.
\end{abstract}


\maketitle

\section{Introduction}
The theory of Brownian motion \cite{mazo02,duplantier05,frey05},
developed in different formulations by Einstein \cite{einstein05},
Smoluchowski \cite{smoluchowski06}
and Langevin \cite{langevin06} around
1905 and 1906, describes
the dynamics of a particle suspended in a fluid.
A prototypical example is a small colloidal object,
e.g.\ a polystyrene bead about a micrometer in size,
floating in water at room temperature.
Even without the action of externally applied forces,
the particle is in an animated and erratic state of motion,
generated on microscopic scales by collisions with the water molecules
and visible on mesoscopic scales as irregular diffusive movement.
Based upon the works mentioned above,
this Brownian motion is most successfully modelled
(on the mesoscopic scales)
by stochastic differential equations \cite{gardiner83,vankampen87,snook07},
augmenting the Newtonian
equations of motion for the particle by the forces from
the surrounding liquid, namely Stokesian
friction proportional to the particle velocity and
thermal fluctuations related to the fluid
temperature.
Both originating from the surrounding
fluid bath as their source, the strengths of
these two forces are related by the fluctuation-dissipation theorem
\cite{einstein05,gardiner83,vankampen87}.

Typically, this set of equations of motion for particle
position and velocity can be simplified by adopting
the so-called overdamped approximation, where one
completely neglects
inertia effects in the particle dynamics. This procedure
is justified because for micrometer-sized
objects suspended in water friction forces are by orders
of magnitude larger than inertial forces \cite{purcell77,dusenbery11},
so that on time- and length-scales accessible under typical
experimental conditions only the overdamped behavior is observable.
Indeed, it takes a delicate experimental effort to actually
measure the velocity of a micrometer-sized Brownian particle
in a liquid \cite{huang11,kheifets14}.

In recent years it has been demonstrated that
the stochastic equations of motion
not only describe
the irregular particle dynamics to high accuracy,
but are also a valid starting point for a
consistent theory of thermodynamic quantities which are
associated with the particle movement,
such as heat, 
work or entropy production
\cite{jarzynski08,jarzynski11,seifert12,vandenBroeck15},
even if the particle is driven away from thermal
equilibrium conditions with the heat bath.
In this newly emerging field called stochastic thermodynamics,
central results of surprising generality have been
discovered in form of fluctuation relations; a recent
summary is provided in \cite{seifert12}.
The entropy production (rate) plays a particularly
important role,
because it constitutes a measure of irreversibility
by relating probabilities of particle trajectories to
their time-reversed counterparts \cite{chetrite08}.

In the present context of a colloidal particle whose
stochastic \emph{dynamics} is well-described by overdamped
equations of motion, the fact that the definition of such a central
concept as entropy production (rate) is based on single
trajectories immediately raises the question of
how neglecting the velocity degrees of freedom may
affect the particle's stochastic \emph{thermodynamics}.
Investigating this question, it has been shown
in \cite{celani12} that the overdamped approximation does
not fully capture the entropy production rate
if the thermal environment is inhomogeneous (but in equilibrium
locally). Rather, there is a contribution to the entropy production
\emph{in addition} to those
predicted from the overdamped approximation,
which can not be obtained from the statistics of the overdamped
trajectories.
Being connected to the breaking of time and velocity
reversal symmetry, 
this phenomenon has been dubbed ``entropic anomaly'' \cite{celani12},
in analogy with similar anomalies encountered in physics.

Perhaps the best known such anomaly is
the viscous dissipative anomaly in turbulence:
the energy dissipation rate in the fluid remains finite
in the limit of vanishing viscosity,
while it vanishes for the viscosity being set to exactly zero \cite{frisch96}.
This anomaly is caused by the breaking of time-reversal symmetry;
time-reversal symmetry is present under
strictly inviscid conditions, but not for finite, yet arbitrarily small viscosity.
Similarly, in quantum physics an anomaly is
related to a symmetry operative on the classical level
(Planck's constant set to zero), which is broken in
the corresponding quantum theory \cite{fujikawa04,camblong01}.

The occurrence of the ``anomalous entropy production''
in the overdamped limit
has been discovered in \cite{celani12} for purely translational motion
of a spherical Brownian particle through an inhomogeneous thermal environment,
see also \cite{spinney12}.
It has been further analyzed in \cite{bo14} and shown to occur in general classes
of stochastic systems, including discrete stochastic processes;
a related analysis of discrete processes is presented in \cite{esposito12}.
The influence of a gradient flow on the ``entropic anomaly''
has been studied in \cite{lan15}, where also effects of particle rotation
have been taken into account. Moreover, the ``entropic anomaly'' has been demonstrated
to affect optimal stochastic transport, i.e.\ driven processes which
optimize entropy production
\cite{bo13}, to induce an efficiency loss
in microscopic stochastic heat-engines \cite{bo13a},
and even to play a non-trivial role in microevolution \cite{bo14a}.
Finally, the ``entropic anomaly'' can be seen as an
explicit example of the general observation that the entropy production
may depend on the scale of description
\cite{esposito12,kawaguchi13,nakayama15}; i.e.\ coarse-graining
the dynamical equations of a physical system by integrating
out a sub-set of degrees of freedom typically reduces entropy
production \cite{kawaguchi13,nakayama15}.
Further concrete systems with such scale-dependent
entropy production are
a harmonic chain of two Brownian particles in contact with
two different heat baths, having finite entropy production which
is reduced to zero when one of the Brownian oscillators is
integrated out \cite{crisanti12}, and
a dimer consisting of two Brownian particles at different temperatures
with a harmonic (but stiff) vs.\ a rigid coupling \cite{chun15}.

In the present paper, we analyze in detail
the Brownian motion of a non-spherical particle,
especially the contributions of rotational
degrees of freedom to the stochastic entropy production and to the
appearance of ``anomalous entropy'' in the overdamped limit. 
We restrict ourselves to the case where there is no
hydrodynamic coupling between rotational and translational degrees
of freedom. Our theory is thus valid for any particle with
three mutually perpendicular symmetry planes \cite{brenner67},
including, in particular, the large class of spheroids and ellipsoids,
but also rods and other rod-like shaped objects.

This paper is organized as follows.
In Sec.~II, we describe the model, provide the fundamental governing equations
for translational and rotational Brownian motion, and introduce the basic
path-wise thermodynamic quantities, such as heat and entropy production,
following the approach of stochastic thermodynamics.
Section III presents the derivation
of the overdamped limit for the concepts introduced in Sec.~II.
Sections IV and V discuss the resulting
overdamped dynamics and overdamped entropy production,
including the anomalous contribution.
In Sec.~VI, the anomalous entropy production is explicitly
calculated for prolate and oblate spheroids,
Sec.~VII treats the case of slightly deformed spherical
particles. We conclude with a short summary and discussion
in Sec.~VIII.
The Appendices A-C contain some additional information
or details of the calculations.

\section{Dynamics and entropy production}
\label{sec:dynamicsEP}
The dynamics of the particle is governed by
external forces and torques, thermal fluctuations, and viscous friction.
Our starting point is to take into account inertia effects as well,
so that we describe such driven Brownian motion by a set of Langevin-Kramers equations
\cite{gardiner83,vankampen87,snook07}
for the particle's translational and rotational degrees of freedom.
The translational motion of the center of mass $\VEC{x}=(x_1,x_2,x_3)$ of the particle
(with mass $m$)
and its velocity $\VEC{v}=(v_1,v_2,v_3)$ is modeled in the laboratory frame,
\begin{subequations}
\label{eq:langevinTrans}
\begin{eqnarray}
\dot{\VEC{x}} & = & \VEC{v} \, ,
\label{eq:xDot}
\\
m\dot{\VEC{v}} & = &
-\gamma\VEC{v} + \VEC{f} + \sqrt{2k_{\mathrm{B}}T}\gamma^{1/2}\VEC{\xi}(t) \, ,
\label{eq:vDot}
\end{eqnarray}
\end{subequations}
where $\VEC{f}=(f_1,f_2,f_3)$ summarizes all deterministic external forces,
$T$ is the temperature ($k_{\mathrm{B}}$ Boltzmann's constant),
and $\VEC{\xi}(t)=(\xi_1(t),\xi_2(t),\xi_3(t))$ are unbiased Gaussian noise
sources with correlations $\langle \xi_i(t)\xi_j(t') \rangle = \delta_{ij}\delta(t-t')$.
Finally, $\gamma$ is the translational viscous friction tensor of the ellipsoid.
This $3\times 3$ tensor is positive definite
and symmetric, so that it has a unique square root,
which we represent by $\gamma^{1/2}$, meaning
$(\gamma^{1/2})\TRANS\gamma^{1/2}=\gamma^{1/2}\gamma^{1/2}=\gamma$
(the symbol $\TRANS$ labels the matrix transpose).

\begin{figure}
\centering
\includegraphics[width=0.4\columnwidth, keepaspectratio=true, angle=0]{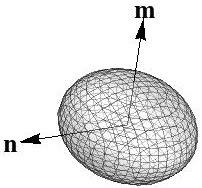}
\caption{Sketch of an ellipsoidal Brownian particle. The two vectors $\VEC{n}$
and $\VEC{m}$ are attached to the particle to parametrize its rotational position,
see also main text, in particular Eqs.~(\ref{eq:nmDot}).}
\label{fig1}
\end{figure}

While the translational dynamics is represented in the
laboratory frame, it turns out to be more convenient to
write the rotational motion in a body-fixed frame (with origin
in the particle's center of mass), because then the inertia
tensor $I$ is independent of particle orientation and thus constant
in time.
The rotational Langevin-Kramers equation for the angular velocity
$\VEC{\omega}=(\omega_1,\omega_2,\omega_3)$
of the ellipsoid is then given by Euler's equation of rigid body
dynamics \cite{goldstein80} with a total torque including not only externally
applied torques but also viscous friction and
thermal noise,
\begin{subequations}
\label{eq:langevinRot}
\begin{equation}
I\dot{\VEC{\omega}}+\VEC{\omega}\times ( I \VEC{\omega})
= -\eta\VEC{\omega} + \VEC{M} + \sqrt{2k_{\mathrm{B}}T}\eta^{1/2}\VEC{\zeta}(t) \, .
\label{eq:omegaDot}
\end{equation}
Here, $\eta$ is the $3\times 3$ symmetric and positive definite
rotational friction tensor with a unique
square root $\eta^{1/2}$,
$\VEC{M}=(M_1,M_2,M_3)$ represents the deterministic torques acting on the
particle,
and $\VEC{\zeta}(t)=(\zeta_1(t),\zeta_2(t),\zeta_3(t))$
are unbiased and delta-correlated Gaussian noise source, which
are independent of the translational ones $\VEC{\xi}(t)$.

In order to specify the orientational position of the
ellipsoid uniquely we choose two orthogonal unit vectors
$\VEC{n}=(n_1,n_2,n_3)$ and $\VEC{m}=(m_1,m_2,m_3)$,
which are rigidly attached to the particle (see Fig.~\ref{fig1}).
Their movement is dictated by the angular velocity
according to the kinematic equations
\begin{equation}
\dot{\VEC{n}} = \VEC{\omega}\times\VEC{n}
\, , \quad
\dot{\VEC{m}} = \VEC{\omega}\times\VEC{m}
\, .
\label{eq:nmDot}
\end{equation} 
\end{subequations}
Since the lengths of these two vectors are set to unity,
and their relative orientation is kept fixed (we choose $\VEC{n}\cdot\VEC{m}=0$),
this representation of the particle orientation has
three free parameters, as expected for a representation
of rotation in three dimensions.
The results we derive in this paper
are of course independent of the specific
representation of the particle rotation. Instead of
(\ref{eq:nmDot}) one could choose a quaternion representation
\cite{altman05}, an Euler angle representation \cite{lan15,goldstein80},
and also a differential geometric representation in terms of
local charts, as discussed in \cite{lan15}.
For the calculation we are going to perform
it turns out, however, that among the global representations the one
in (\ref{eq:nmDot}) is the most convenient.

We emphasize again that translational motion (\ref{eq:langevinTrans})
is written in the laboratory frame, while rotational dynamics
(\ref{eq:langevinRot}) is given in a reference frame fixed to
the particle with the origin being located
at the center of mass of the particle.
We do not
distinguish quantities in the different reference frames by explicit labels though,
in order to keep notation as simple as possible.
In other words, throughout the present manuscript we follow the rule
that all quantities associated with translation are represented in
the laboratory system, while quantities associated with rotation
are represented in the body fixed coordinate system.

A central property
of our setup is the presence of an inhomogeneous thermal environment,
i.e.\ we allow the temperature $T$ to depend on space and time,
but with thermal equilibrium being valid locally.
Likewise, the friction tensors $\gamma$ and $\eta$ are assumed to be
functions of space and time, for instance because the fluid viscosity changes
with the spatial variations of temperature or due to hydrodynamic
effects close to boundaries \cite{happel83}.
Moreover, since translation is considered in the laboratory
frame, the translational friction tensor depends on
the particle orientation, while the body-fixed rotational friction
tensor does not. Concerning the external deterministic
forces and torques, we allow for the most general case
and take into account variations on position, orientation and time.
Therefore, we have the following functional dependencies for
the quantities appearing in (\ref{eq:langevinTrans}), (\ref{eq:langevinRot}):
\begin{subequations}
\label{eq:dep}
\begin{eqnarray}
T       & = & T(\VEC{x},t) \, , \label{eq:depT}\\
\gamma  & = & \gamma(\VEC{x},\VEC{n},\VEC{m},t) \, , \\
\eta    & = & \eta(\VEC{x},t) \, , \\
\VEC{f} & = & \VEC{f}(\VEC{x},\VEC{n},\VEC{m},t) \, , \\
\VEC{M} & = & \VEC{M}(\VEC{x},\VEC{n},\VEC{m},t) \, .
\end{eqnarray}
\end{subequations}
%
%

The equations of motion (\ref{eq:langevinTrans}), (\ref{eq:langevinRot})
generate, for given initial position and orientation of the ellipsoid,
stochastic trajectories in translational and orientational configuration
space.
According to stochastic thermodynamics \cite{jarzynski11,seifert12},
an entropy production $\Delta S$ is associated with such stochastic
trajectories, which involves contributions from the change
of particle entropy $\Delta S_{\mathrm{p}}$ and from the entropy production
in the environment $\Delta S_{\mathrm{env}}$,
\begin{equation}
\label{eq:DeltaS}
\Delta S = \Delta S_{\mathrm{p}} + \Delta S_{\mathrm{env}} \, .
\end{equation}
We briefly summarize the well-known results for translational Brownian motion
\cite{seifert05}
and extend them to include rotational stochastic dynamics.

The particle entropy is defined as the state function \cite{seifert05}
\begin{equation}
\label{eq:Sp}
S_{\mathrm{p}} = -k_{\mathrm{B}} \, \ln p(\VEC{x},\VEC{v},\VEC{n},\VEC{m},\VEC{\omega},t) \, ,
\end{equation}
where $p(\VEC{x},\VEC{v},\VEC{n},\VEC{m},\VEC{\omega},t)$ is the solution of the
Fokker-Planck equation associated with (\ref{eq:langevinTrans}), (\ref{eq:langevinRot}).
In our case, $S_{\mathrm{p}}$ does not only depend on translational degrees of freedom
but also includes particle orientation, parametrized by $\VEC{n}$, $\VEC{m}$ and $\VEC{\omega}$.
The change of particle entropy along a trajectory which starts
at a point $(\VEC{x}_0,\VEC{v}_0,\VEC{n}_0,\VEC{m}_0,\VEC{\omega}_0)$ in configuration space at time
$t_0$ and is located at a point $(\VEC{x}(t),\VEC{v}(t),\VEC{n}(t),\VEC{m}(t),\VEC{\omega}(t))$ at a later time $t$
is given by
\begin{multline}
\label{eq:DeltaSp}
\Delta S_{\mathrm{p}} = k_{\mathrm{B}} \, \ln p(\VEC{x}_0,\VEC{v}_0,\VEC{n}_0,\VEC{m}_0,\VEC{\omega}_0,t_0)
\\
- k_{\mathrm{B}} \, \ln p(\VEC{x}(t),\VEC{v}(t),\VEC{n}(t),\VEC{m}(t),\VEC{\omega}(t),t) \, .
\end{multline}

The entropy production in the environment $\mathrm{d}S_{\mathrm{env}}$ for an
infinitesimal displacement of the particle is related to the
heat $\delta Q$ released to the thermal bath during this displacement,
\begin{equation}
\label{eq:dSenv}
\mathrm{d}S_{\mathrm{env}} = \frac{\delta Q}{T} \, .
\end{equation}
Following Sekimoto's stochastic energetics approach \cite{sekimoto10},
the heat is identified with the work done by the particle on the
thermal environment
\cite{Note1}.
This work results from the forces the particle
exerts on the environment during its movement as reaction forces
to viscous friction and thermal fluctuations, such that
\begin{eqnarray}
\delta Q  & = & -[-\gamma\VEC{v} + \sqrt{2k_{\mathrm{B}}T}\gamma^{1/2}\VEC{\xi}(t)] \circ \VEC{v}\,\mathrm{d}t
\nonumber \\
&&
-[-\eta\VEC{\omega} + \sqrt{2k_{\mathrm{B}}T}\eta^{1/2}\VEC{\zeta}(t)] \circ \VEC{\omega}\,\mathrm{d}t
\label{eq:deltaQ}
\end{eqnarray}
for a total particle displacement consisting of a translational
increment $\mathrm{d}\VEC{x}=\VEC{v}\mathrm{d}t$ and a rotational
increment $\VEC{\omega}\mathrm{d}t$ during the time step $\mathrm{d}t$.
The symbol $\circ$ denotes the scalar product evaluated according
to the Stratonovich rule (midpoint regularization) \cite{sekimoto10}.
Note that the heat $\delta Q$ as well as its two additive contributions
from translation and rotation are scalar quantities and thus
invariant under rotation of reference frame, a property which we exploited
in (\ref{eq:deltaQ}) by representing the translational part
in laboratory coordinates and the rotational part in the body-fixed frame.
The entropy produced by the particle in the environment along a
stochastic trajectory is the integral of (\ref{eq:dSenv}) over that path
with $\delta Q$ given by (\ref{eq:deltaQ}). Using (\ref{eq:vDot})
and (\ref{eq:omegaDot}), and the identity
$[\VEC{\omega} \times (I \cdot \VEC{\omega})] \cdot \VEC{\omega}=0$
we find
\begin{multline}
\label{eq:DeltaSenv}
\Delta S_{\mathrm{env}} = \int_{t_0}^{t}
\frac{1}{T}
\left[
\VEC{f}\cdot\VEC{v}\,\mathrm{d}t' + \VEC{M}\cdot\VEC{\omega}\,\mathrm{d}t'
\right.
\\
\left. 
- m\VEC{v}\circ\mathrm{d}\VEC{v}(t') - (I\VEC{\omega})\circ\mathrm{d}\VEC{\omega}(t')
\right]
\, .
\end{multline}

In Appendix \ref{app:EP} we verify that this entropy production
in the environment can be interpreted
as a measure of irreversibility \cite{chetrite08,vandenBroeck15},
by relating it to the ratio of
probabilities of forward and time-reversed paths 
for the combined translational and rotational motion.
We also verify that the total entropy production $\Delta S$,
given by (\ref{eq:DeltaS}) with (\ref{eq:DeltaSp}), (\ref{eq:DeltaSenv}),
fulfills the usual fluctuation theorem \cite{seifert05},
\begin{equation}
\label{eq:FT}
\left\langle e^{-\Delta S/k_{\mathrm{B}}} \right\rangle = 1 \, ,
\end{equation}
when averaged over the path
ensemble with fixed initial configuration,
denoted by $\langle \ldots \rangle$.
Its average is therefore always positive
(by Jensen's inequality)
\begin{equation}
\left\langle \Delta S \right\rangle \geq 0 \, ,
\end{equation}
in accordance with the second law.

\section{Overdamped limit}
As described in the Introduction, the typical system we want to study
with our model (\ref{eq:langevinTrans}), (\ref{eq:langevinRot})
is a colloidal ellipsoidal particle of micrometer-size (or nanometer-size)
suspended in water with an inhomogeneous temperature distribution
around room temperature. It is well known that in such
systems the effects of the fluid bath in form of thermal fluctuations
and viscous friction have significant influence on the particle
dynamics.
Indeed, viscous friction effects are by orders of magnitude larger than
inertia effects \cite{purcell77,dusenbery11},
so that one typically disregards inertia
completely and adopts the so-called \emph{overdamped approximation}
by simply putting mass equal to zero, $m=0$.
In a strict mathematical sense, however, this is
generally not a
valid procedure as it changes the order of the
differential equations of motion (\ref{eq:langevinTrans}),
(\ref{eq:langevinRot}), so that small inertia effects
(as compared to friction forces)
actually correspond to a singular perturbation.
Therefore, the \emph{overdamped limit} should be performed
with some care in a genuine perturbative way.
Although the overdamped approximation and the overdamped
limit give identical results in many cases, this can
not be taken for granted,
in particular for stochastic differential
equations like (\ref{eq:langevinTrans}), (\ref{eq:langevinRot})
\cite{ryter81,sancho82,jayannavar95,hottovy12,yang13,widder89,freidlin04,lau07},
involving heterogeneous heat baths,
or for functionals along stochastic trajectories generated by
(\ref{eq:langevinTrans}), (\ref{eq:langevinRot}), like
heat \cite{celani12,matsuo00} or the
entropy production (\ref{eq:DeltaSenv})
\cite{celani12,bo14,lan15}.

\subsection{Multiple time-scales}
\label{sec:timeScales}
A standard tool for the rigorous derivation of singular limits
is the so-called multiple time-scale technique \cite{bender99},
which exploits that the singular limit is tied to the appearance
of various well-separated dynamical time-scales in the system.
In our case, the dominance of viscous friction over inertia effects is
related to the existence of two distinct time-scales
in the equations of motion (\ref{eq:langevinTrans}), (\ref{eq:langevinRot}).
The first one is the time over which the velocity
degrees of freedom relax and reach their stationary distribution. It is given
by
\begin{equation}
\label{eq:tauv}
\tau_v = \frac{m}{\gamma_0} \, ,
\end{equation}
where $\gamma_0$ is the ``typical'' friction
coefficient of the particle, e.g., for a spherical particle of radius $a$
it corresponds to Stokes friction $\gamma_0 = 6\pi \nu a$ ($\nu$ is the
dynamic viscosity of the medium the particle is suspended in).
For a micrometer-sized particle in water, $\tau_v$ is of the order of microseconds
\cite{purcell77}. The second time-scale is the time after which
one can observe the diffusive motion of the particle over a
detectable distance, typically about the particle size $a$,
\begin{equation}
\label{eq:taux}
\tau_x = \frac{a^2\gamma_0}{k_{\mathrm{B}}T_0} \, ,
\end{equation}
with $T_0$ being the
average temperature of the bath.
Under identical conditions, i.e.\ 
for a micrometer-sized particle in water, $\tau_x$ is of the order of seconds.
Note that although we have based our
definitions of $\tau_v$ and $\tau_x$ on translational motion, the
time-scales for rotation are basically the same, because
the ``rotational mass'' (i.e.\ moment of inertia) scales as
$I \sim m a^2$ and the rotational viscous friction as $\eta \sim \gamma a^2$.

Based on this separation of time-scales, we
calculate the overdamped limit 
of the entropy production (\ref{eq:DeltaS})
[with (\ref{eq:DeltaSp}), (\ref{eq:DeltaSenv})] using
the multiple time-scale method.
It is convenient to perform this limit
for the generating function of the entropy production
in the environment (\ref{eq:DeltaSenv}) \cite{celani12}.
As side results, we then find
the well-known overdamped
versions of the equations of
motion (\ref{eq:langevinTrans}), (\ref{eq:langevinRot}),
as well as the relation connecting the probability density
$p$ in (\ref{eq:Sp}) with its overdamped counterpart.
The latter is needed to deduce the overdamped limit of
the change in particle entropy (\ref{eq:DeltaSp}),
and from that the full entropy production (\ref{eq:DeltaS}).

\subsection{Generating function}
In order to define the generating function of the 
entropy production in the environment $\Delta S_{\mathrm{env}}$,
we first rewrite (\ref{eq:DeltaSenv}) as an integral in
$\mathrm{d}t'$ only.
Observing that
$k_{\mathrm{B}}\,\mathrm{d}[{(I\VEC{\omega})\cdot\VEC{\omega}}/{2k_{\mathrm{B}}T}]
= k_{\mathrm{B}}\,\mathrm{d}[{\VEC{\omega}\TRANS I \VEC{\omega}}/{2k_{\mathrm{B}}T}]
= (I\VEC{\omega})\circ\mathrm{d}\VEC{\omega}/T - [{\VEC{\omega}\TRANS I \VEC{\omega}}/{2T^2}]\,\mathrm{d}T$
and that also
$k_{\mathrm{B}}\,\mathrm{d}[{\VEC{\omega}\TRANS I \VEC{\omega}}/{2k_{\mathrm{B}}T}]
= -k_{\mathrm{B}}\,\mathrm{d}\ln w_\omega - (3k_{\mathrm{B}}/2T)\,\mathrm{d}T$
with the Maxwell-Boltzmann distribution for the angular velocity
\begin{subequations}
\label{eq:w}
\begin{equation}
\label{eq:w_omega}
w_\omega = \frac{\sqrt{\det(I)}}{(2\pi k_{\mathrm{B}} T)^{3/2}} \exp \left[ -\frac{\VEC{\omega}\TRANS I \VEC{\omega}}{2k_{\mathrm{B}}T} \right] \, ,
\end{equation}
we may write $-(I\VEC{\omega})\circ\mathrm{d}\VEC{\omega}/T$
in (\ref{eq:DeltaSenv}) as
$k_{\mathrm{B}}\,\mathrm{d}\ln w_\omega + [(3k_{\mathrm{B}}T - {\VEC{\omega}\TRANS I \VEC{\omega}})/{2T^2}]\,\mathrm{d}T$.
The term $-m\VEC{v}\circ\mathrm{d}\VEC{v}$
can be recast in a similar way as
$k_{\mathrm{B}}\,\mathrm{d}\ln w_v + [(3k_{\mathrm{B}}T - m\VEC{v}^2)/{2T^2}]\,\mathrm{d}T$
by making use of the
Maxwell-Boltmann distribution for the translational velocity,
\begin{equation}
\label{eq:w_v}
w_v = \left( \frac{m}{2\pi k_{\mathrm{B}} T} \right)^{3/2} \exp \left[ -\frac{m \VEC{v}^2}{2k_{\mathrm{B}}T} \right] \, .
\end{equation}
\end{subequations}
The $\mathrm{d}T$ contributions are a consequence of the
inhomogeneity of the thermal environment
leading to temperature variations along the particle
trajectory, which are given as 
$\mathrm{d}T = (\partial T/\partial t + \VEC{v}\cdot\partial T/\partial\VEC{x})\,\mathrm{d}t$
according to (\ref{eq:depT}),
where $\partial T/\partial\VEC{x} = (\partial T/\partial x_1,\partial T/\partial x_2,\partial T/\partial x_3)$.

Collecting all these pieces together, we
rearrange the resulting stochastic integral
into three parts \cite{celani12},
$\Delta S_{\mathrm{reg}}$, $\Delta S_{\mathrm{time}}$ and $\Delta S_{\mathrm{anom}}$,
related to the regular entropy production,
an entropy production due to time-changes of $T$,
and a part of the entropy production
related to spatial temperature variations,
which we show to
yield an anomalous contribution in the overdamped limit.
The final result for $\Delta S_{\mathrm{env}}$ from (\ref{eq:DeltaSenv})
thus reads
\begin{multline}
\label{eq:DeltaSenv_split}
\Delta S_{\mathrm{env}} = 
-k_{\mathrm{B}} \ln w_v(t_0)w_\omega(t_0) + k_{\mathrm{B}} \ln w_v(t)w_\omega(t)
\\
+ \Delta S_{\mathrm{reg}} + \Delta S_{\mathrm{time}} + \Delta S_{\mathrm{anom}}
\end{multline}
with
\begin{widetext}
\begin{subequations}
\begin{eqnarray}
\Delta S_{\mathrm{reg}} & = &
\int_{t_0}^{t} \left( \frac{\VEC{f}\cdot\VEC{v}}{T} - \frac{\VEC{v}}{T}\cdot\frac{\partial k_{\mathrm{B}}T}{\partial\VEC{x}} + \frac{\VEC{M}\cdot\VEC{\omega}}{T} \right) \mathrm{d}t'
\, ,
\label{eq:DeltaSreg} \\[2ex]
\Delta S_{\mathrm{time}} & = & 
\int_{t_0}^{t} \left( \frac{3k_{\mathrm{B}}T - m\VEC{v}^2}{2T^2} + \frac{3k_{\mathrm{B}}T - \VEC{\omega}\TRANS I \VEC{\omega}}{2T^2} \right)
\frac{\partial T}{\partial t} \,\mathrm{d}t'
\, ,
\label{eq:DeltaStime} \\[2ex]
\Delta S_{\mathrm{anom}} & = &
\int_{t_0}^{t} \left( \frac{5k_{\mathrm{B}}T - m\VEC{v}^2}{2T^2} + \frac{3k_{\mathrm{B}}T - \VEC{\omega}\TRANS I \VEC{\omega}}{2T^2} \right)
\VEC{v}\cdot\frac{\partial T}{\partial\VEC{x}} \,\mathrm{d}t'
\, ,
\label{eq:DeltaSanom}
\end{eqnarray}
\label{eq:DeltaS3}
\end{subequations}
\end{widetext}

By splitting the entropy
production into these three contributions
we can essentially separate the effects
of time- and spatial variations of temperature,
(\ref{eq:DeltaStime}) and (\ref{eq:DeltaSanom}),
from the usual ``regular'' contribution (\ref{eq:DeltaSreg}).
We point out though that the regular part (\ref{eq:DeltaSreg}),
contains a temperature-gradient term,
which is compensated by the factor 5/2 (instead of 3/2)
in the first term of (\ref{eq:DeltaSanom}).
The appearance of the temperature-gradient
is inspired by the (\textit{a posteriori}) observation
that the entropy production in the overdamped approximation,
when inertia effects are simply disregarded,
exactly corresponds to the expression (\ref{eq:DeltaSreg}).
This can be verified by calculating path probability ratios,
see \cite{chetrite08} and the Supplementary Material of \cite{celani12}.

The joint generating function of the three
contributions (\ref{eq:DeltaS3}) to the
entropy production is \cite{celani12}
\begin{widetext}
\begin{multline}
\label{eq:Gs}
G_{s_1 s_2 s_3}(\VEC{x},\VEC{v},\VEC{n},\VEC{m},\VEC{\omega},t|\VEC{x}_0,\VEC{v}_0,\VEC{n}_0,\VEC{m}_0,\VEC{\omega}_0,t_0)
= \left\langle \exp(-s_1\Delta S_{\mathrm{reg}} - s_2\Delta S_{\mathrm{time}} - s_3 \Delta S_{\mathrm{anom}})
\right. \\ \left.
\delta(\VEC{x}(t)-\VEC{x})\delta(\VEC{v}(t)-\VEC{v})\delta(\VEC{n}(t)-\VEC{n})\delta(\VEC{m}(t)-\VEC{m})\delta(\VEC{\omega}(t)-\VEC{\omega}) \right\rangle
\, ,
\end{multline}
where the average is taken over paths with fixed initial conditions
$\VEC{x}_0$, $\VEC{v}_0$, $\VEC{n}_0$, $\VEC{m}_0$, $\VEC{\omega}_0$ at time $t_0$, as before.
It can be shown to obey the forward Feynman-Kac formula \cite{mazo02,celani12}
\begin{multline}
\label{eq:FK}
\frac{\partial G_{s_1 s_2 s_3}}{\partial t} - \OP{A}^\dagger G_{s_1 s_2 s_3}
= -\left[
s_1 \left( \frac{\VEC{f}\cdot\VEC{v}}{T} - \frac{\VEC{v}}{T}\cdot\frac{\partial k_{\mathrm{B}}T}{\partial\VEC{x}} + \frac{\VEC{M}\cdot\VEC{\omega}}{T} \right)
+ s_2 \left( \frac{3k_{\mathrm{B}}T - m\VEC{v}^2}{2T^2} + \frac{3k_{\mathrm{B}}T - \VEC{\omega}\TRANS I \VEC{\omega}}{2T^2} \right)
\frac{\partial T}{\partial t}
\right. \\ \left.
+ s_3 \left( \frac{5k_{\mathrm{B}}T - m\VEC{v}^2}{2T^2} + \frac{3k_{\mathrm{B}}T - \VEC{\omega}\TRANS I \VEC{\omega}}{2T^2} \right)
\VEC{v}\cdot\frac{\partial T}{\partial\VEC{x}}
\right] G_{s_1 s_2 s_3}
\, .
\end{multline}
Note that for $s_1=0$, $s_2=0$, and $s_3=0$, the generating function represents
the probability density $p$ used in (\ref{eq:Sp}),
\begin{eqnarray}
G_{000}(\VEC{x},\VEC{v},\VEC{n},\VEC{m},\VEC{\omega},t|\VEC{x}_0,\VEC{v}_0,\VEC{n}_0,\VEC{m}_0,\VEC{\omega}_0,t_0)& = & \left\langle \delta(\VEC{x}(t)-\VEC{x})\delta(\VEC{v}(t)-\VEC{v})\delta(\VEC{n}(t)-\VEC{n})\delta(\VEC{m}(t)-\VEC{m})\delta(\VEC{\omega}(t)-\VEC{\omega}) \right\rangle
\nonumber\\
& = & p(\VEC{x},\VEC{v},\VEC{n},\VEC{m},\VEC{\omega},t) \, .
\label{eq:G000}
\end{eqnarray}
\end{widetext}
Accordingly, the operator $\OP{A}^\dagger$ is the generator of the
combined diffusion process for translation and rotation associated with
(\ref{eq:langevinTrans}), (\ref{eq:langevinRot});
its specific expression is given below in Eq.~(\ref{eq:Atilde})
(in dimensionless form).

\subsection{Dimensionless representation}
Since inertial effects in (\ref{eq:langevinTrans}), (\ref{eq:langevinRot})
are orders of magnitude smaller than friction and other forces, the
various terms in (\ref{eq:FK}) may be of considerably different magnitude as well.
For a detailed analysis, we rewrite all
quantities appearing in (\ref{eq:FK}) by
introducing dimensionless representations of order one, so that
the different magnitudes of terms show up as dimensionless
small (or large) prefactors, which we expect to be related to
the ratio of the two distinct time-scales $\tau_v$ and $\tau_x$
(see (\ref{eq:tauv}) and (\ref{eq:taux})).
Our choice 
is guided by physical intuition and by the characteristics of 
the systems we intend to model with (\ref{eq:langevinTrans}), (\ref{eq:langevinRot}).
Most importantly, due to the separation of time-scales $\tau_v \ll \tau_x$,
we expect the (translational and rotational)
velocity degrees of freedom to equilibrate ``instantaneously''
and become of the order the thermal velocity.
In contrast, positional degrees of freedom
change significantly only on the ``large'' scales
$a$ and $\tau_x$, so that we measure length and time
using these units.
External forces and torques are
assumed to be of about the same size as the thermal
fluctuating forces. We therefore make the following ansatz for
relating dimensionfull and dimensionless quantities
(denoted by a tilde):
\begin{subequations}
\label{eq:dimfull2dimless}
\begin{align}
t &= \tau_x\tilde{t} \, , \\
\VEC{v} &= \sqrt{\frac{k_{\mathrm{B}}T_0}{m}}\tilde{\VEC{v}} \, ,	& 	\VEC{\omega} &= \sqrt{\frac{k_{\mathrm{B}}T_0}{ma^2}}\tilde{\VEC{\omega}} \, , \\
\VEC{x} &= a\tilde{\VEC{x}} \, ,	 								& 	\VEC{n} &= \tilde{\VEC{n}} \, , \; \VEC{m} = \tilde{\VEC{m}} , \\
\VEC{f} &= \frac{k_{\mathrm{B}}T_0}{a}\tilde{\VEC{f}} \, ,			&	\VEC{M} &= k_{\mathrm{B}}T_0 \tilde{\VEC{M}} \, .
\end{align}
We furthermore express $I$ in terms of $m$ and the length-scale $a$,
\begin{equation}
I = ma^2 \tilde{I} \, ,
\end{equation}
the friction tensors in terms of the ``typical'' friction coefficient $\gamma_0$,
\begin{equation}
\gamma = \gamma_0 \tilde{\gamma} \, ,	\quad	\eta = \gamma_0 a^2 \tilde{\eta} \, ,
\end{equation}
and the temperature field by the average temperature $T_0$,
\begin{equation}
k_{\mathrm{B}}T = k_{\mathrm{B}}T_0 \, \tilde{T} \, .
\end{equation}
\end{subequations}

Plugging the relations (\ref{eq:dimfull2dimless}) into (\ref{eq:FK})
and defining the dimensionless variable $\tilde{s_i} = k_{\mathrm{B}} s_i$,
we obtain the dimensionless form of the forward Feynman-Kac equation
\begin{equation}
\label{eq:FKdimless}
\left( \frac{\partial}{\partial\tilde t} - \EPS^{-1}\tilde{\OP{L}}^\dagger - \EPS^{-2}\tilde{\OP{M}}^\dagger - \tilde{\OP{N}}^\dagger \right)
G_{\tilde{s}_1 \tilde{s}_2 \tilde{s}_3} = 0 \, ,
\end{equation}
with
\begin{widetext}
\begin{subequations}
\begin{align}
\tilde{\OP{L}}^\dagger & = 
-\tilde{v}_i \frac{\partial}{\partial\tilde{x}_i} - \tilde{f}_i \frac{\partial}{\partial\tilde{v}_i}
-\epsilon_{ijk}\tilde{\omega}_j \left( \frac{\partial}{\partial\tilde{n}_i}\tilde{n}_k + \frac{\partial}{\partial\tilde{m}_i}\tilde{m}_k \right)
-(\tilde{I}^{-1})_{ij}\tilde{M}_j \frac{\partial}{\partial\tilde{\omega}_i}
+(\tilde{I}^{-1})_{ij}\tilde{I}_{lm}\epsilon_{jkl}\frac{\partial}{\partial\tilde{\omega}_i}\tilde{\omega}_k \tilde{\omega}_m
\nonumber\\
&\qquad
-\tilde{s}_1 \left[ \frac{\tilde{f}_i\tilde{v}_i}{\tilde{T}}
				  - \frac{\tilde{v}_i}{\tilde{T}}\left(\frac{\partial\tilde{T}}{\partial\tilde{x}_i}\right)
				  + \frac{\tilde{M}_i\tilde{\omega}_i}{\tilde{T}} \right]
-\tilde{s}_3 \left( \frac{5\tilde{T}\tilde{v}_i-\tilde{v}_j\tilde{v}_j\tilde{v}_i}{2\tilde{T}^2}
				  + \frac{3\tilde{T}\tilde{v}_i-\tilde{I}_{jk}\tilde{\omega}_j\tilde{\omega}_k\tilde{v}_i}{2\tilde{T}^2} \right)
			 \left( \frac{\partial\tilde{T}}{\partial\tilde{x}_i} \right) \, ,
\label{eq:Ldagger} \\
\tilde{\OP{M}}^\dagger & =
\tilde{\gamma}_{ij}\frac{\partial}{\partial\tilde{v}_i}\tilde{v}_j
+ \tilde{T}\tilde{\gamma}_{ij}\frac{\partial}{\partial\tilde{v}_i}\frac{\partial}{\partial\tilde{v}_j}
+ (\tilde{I}^{-1})_{ij}\tilde{\eta}_{jk}\frac{\partial}{\partial\tilde{\omega}_i}\tilde{\omega}_k
+ \tilde{T}(\tilde{I}^{-1})_{il}(\tilde{I}^{-1})_{kj}\tilde{\eta}_{lk}\frac{\partial}{\partial\tilde{\omega}_i}\frac{\partial}{\partial\tilde{\omega}_j} \, ,
\label{eq:Mdagger} \\
\tilde{\OP{N}}^\dagger & =
-\tilde{s}_2 \left( \frac{3\tilde{T}-\tilde{v}_i\tilde{v}_i}{2\tilde{T}^2}
	 			  + \frac{3\tilde{T}-\tilde{I}_{ij}\tilde{\omega}_i\tilde{\omega}_j}{2\tilde{T}^2} \right)
			 \left( \frac{\partial\tilde{T}}{\partial\tilde t} \right) \, ,
\label{eq:Ndagger}
\end{align}
\end{subequations}
\end{widetext}
where we switched to index notation for convenience
with summation over repeated indices being understood.
In (\ref{eq:FKdimless}), we defined
\begin{equation}
\label{eq:eps}
\EPS = \sqrt{\frac{\tau_v}{\tau_x}} \ll 1
\end{equation}
as small parameter, expressing the time-scale separation in the system.
Note that the dimensionless version $\tilde{\OP{A}}^\dagger$ of
the generator $\OP{A}^\dagger$
of the diffusion process used in (\ref{eq:FK}) is given as
\begin{eqnarray}
\tilde{\OP{A}}^\dagger & = & \left[
	\EPS^{-1}\tilde{\OP{L}}^\dagger + \EPS^{-2}\tilde{\OP{M}}^\dagger + \tilde{\OP{N}}^\dagger
	\right]_{\tilde{s}_1=0,\tilde{s}_2=0,\tilde{s}_3=0}
\nonumber \\
& = & \left. \EPS^{-1}\tilde{\OP{L}}^\dagger \right|_{\tilde{s}_1=0,\tilde{s}_3=0} + \EPS^{-2}\tilde{\OP{M}}^\dagger \, .
\label{eq:Atilde}
\end{eqnarray}

\subsection{Perturbation expansion}
The dimensionless equation of motion (\ref{eq:FKdimless})
for the generating function (\ref{eq:Gs})
constitutes the starting point of our analysis,
with the goal to derive its overdamped counterpart
in the (singular) limit $\EPS=\sqrt{\tau_v/\tau_x} \to 0$.
To explicitly account for the observation that
the system exhibits dynamics on different time-scales
we apply the following multi-scale procedure.
First, we introduce
time variables $\theta$ and $\tau$ corresponding to the scales given
by $\tau_v$ and $\tau_x$ (see (\ref{eq:tauv}) and (\ref{eq:taux})), 
and a variable $\vartheta$ for the intermediate scale
\cite{celani12,bo14,pavliotis08},
\begin{equation}
\label{eq:times}
\theta = \EPS^{-2}\tilde{t}
\, , \quad
\vartheta = \EPS^{-1}\tilde{t}
\, , \quad
\tau = \tilde{t}
\, .
\end{equation}
We assume that the external time-changes of temperature, friction coefficients,
forces and torques occur on the slow time-scale $\tau$ only,
so that the dimensionless versions of (\ref{eq:dep}) read
\begin{subequations}
\label{eq:depdimless}
\begin{eqnarray}
\tilde{T}       & = & \tilde{T}(\tilde{\VEC{x}},\tau) \, , \\
\tilde{\gamma}  & = & \tilde{\gamma}(\tilde{\VEC{x}},\tilde{\VEC{n}},\tilde{\VEC{m}},\tau) \, , \\
\tilde{\eta}    & = & \tilde{\eta}(\tilde{\VEC{x}},\tau) \, , \\
\tilde{\VEC{f}} & = & \tilde{\VEC{f}}(\tilde{\VEC{x}},\tilde{\VEC{n}},\tilde{\VEC{m}},\tau) \, , \\
\tilde{\VEC{M}} & = & \tilde{\VEC{M}}(\tilde{\VEC{x}},\tilde{\VEC{n}},\tilde{\VEC{m}},\tau) \, .
\end{eqnarray}
\end{subequations}
Then, we expand $G_{\tilde{s}_1\tilde{s}_2\tilde{s}_3}$ in powers of $\EPS$,
\begin{equation}
\label{eq:expansionG}
G_{\tilde{s}_1\tilde{s}_2\tilde{s}_3} = G^{(0)} + \EPS G^{(1)} + \EPS^2 G^{(2)} + \ldots \, ,
\end{equation}
where all the $G^{(i)}$ are \textit{a priori} assumed to be functions of all variables
$\tilde{\VEC{x}}$, $\tilde{\VEC{v}}$, $\tilde{\VEC{n}}$, $\tilde{\VEC{m}}$, $\tilde{\VEC{\omega}}$
and of all the three different times $\theta$, $\vartheta$, $\tau$ defined in
(\ref{eq:times})
(of course they also depend on the parameters $\tilde{s}_1$,
$\tilde{s}_2$, $\tilde{s}_3$, but we skip the corresponding subscripts for
notational simplicity).
As a consequence, the time-derivative in (\ref{eq:FKdimless})
turns into
\begin{equation}
\label{eq:Dtimes}
\frac{\partial}{\partial\tilde{t}}
= \EPS^{-2}\frac{\partial}{\partial\theta} + \EPS^{-1}\frac{\partial}{\partial\vartheta} + \frac{\partial}{\partial\tau} \, .
\end{equation}

Inserting (\ref{eq:expansionG}), (\ref{eq:Dtimes}) into (\ref{eq:FKdimless})
and equating terms of equal power in $\EPS$, we find a hierarchy of coupled
equations with lowest order (order $\EPS^{-2}$)
\begin{equation}
\frac{\partial G^{(0)}}{\partial\theta}-\tilde{\OP{M}}^\dagger G^{(0)} = 0 \, .
\end{equation}
From (\ref{eq:Mdagger}) we see that $\tilde{\OP{M}}^\dagger$ is an operator in the
fast degrees of freedom $\VEC{v}$, $\VEC{\omega}$ only, so that their dynamics
is indeed tied to the fast time $\theta$, as expected.
Since there are no explicit $\theta$ dependences in $\tilde{\OP{M}}^\dagger$
(see (\ref{eq:depdimless})),
and since we are not interested in the relaxation processes
of the fast degrees of freedom on $\theta$ time-scales, we can set 
$\partial G^{(i)}/\partial\theta = 0$ (for all $i$) in the following.
Then, the first three in the hierarchy of equations read
\begin{subequations}
\label{eq:hierarchy}
\begin{align}
\tilde{\OP{M}}^\dagger G^{(0)} & = 0
\, , \label{eq:order-2} \\
\tilde{\OP{M}}^\dagger G^{(1)} & =
\frac{\partial G^{(0)}}{\partial\vartheta} - \tilde{\OP{L}}^\dagger G^{(0)}
\, , \label{eq:order-1} \\
\tilde{\OP{M}}^\dagger G^{(2)} & =
\frac{\partial G^{(0)}}{\partial\tau} - \tilde{\OP{N}}^\dagger G^{(0)} + \frac{\partial G^{(1)}}{\partial\vartheta} - \tilde{\OP{L}}^\dagger G^{(1)}
\, , \label{eq:order0}
\end{align}
\end{subequations}
with the first line collecting order $\EPS^{-2}$ terms,
the second line order $\EPS^{-1}$ terms
and the third line order $\EPS^0$ terms.

The solution to (\ref{eq:order-2}) is
\begin{equation}
\label{eq:G0}
G^{(0)} = g^{(0)}(\tilde{\VEC{x}},\tilde{\VEC{n}},\tilde{\VEC{m}},\vartheta,\tau)
\tilde{w}_v(\tilde{\VEC{v}}|\tilde{\VEC{x}},\tau)
\tilde{w}_\omega(\tilde{\VEC{\omega}}|\tilde{\VEC{x}},\tau) \, ,
\end{equation}
where $\tilde{w}_v(\tilde{\VEC{v}}|\tilde{\VEC{x}},\tau)$ and
$\tilde{w}_\omega(\tilde{\VEC{\omega}}|\tilde{\VEC{x}},\tau)$
are the dimensionless counterparts of the Maxwell-Boltzmann
distributions for translational and rotational velocity
(at given position
and time) defined in (\ref{eq:w})
\cite{Note2},
and where $g^{(0)}(\tilde{\VEC{x}},\tilde{\VEC{n}},\tilde{\VEC{m}},\vartheta,\tau)$
is an unknown function of only the slow degrees of freedom
according to the general definition
\begin{equation}
g^{(i)}(\tilde{\VEC{x}},\tilde{\VEC{n}},\tilde{\VEC{m}},\vartheta,\tau)
= \int \mathrm{d}\tilde{\VEC{v}}\mathrm{d}\tilde{\VEC{\omega}} \, G^{(i)} \, .
\end{equation}

To proceed with the higher order equations
(\ref{eq:order-1}), (\ref{eq:order0}) we employ the solvability
condition. It states that the inhomogeneities on the right-hand
sides need to be orthogonal to the nullspace of the
operator $\tilde{\OP{M}}$ adjoint to $\tilde{\OP{M}}^\dagger$
(Fredholm alternative, see, e.g., \cite{pavliotis08}).
As we can see
from (\ref{eq:Mdagger}), the nullspace of $\tilde{\OP{M}}$
contains the functions which are constant in
$\tilde{\VEC{v}}$ and $\tilde{\VEC{\omega}}$.
Therefore, the solvability condition for (\ref{eq:order-1}) reads
$\int\mathrm{d}\tilde{\VEC{v}}\mathrm{d}\tilde{\VEC{\omega}}\,(\partial G^{(0)}/\partial\vartheta - \tilde{\OP{L}}^\dagger G^{(0)})=0$.
It is straightforward to show from the explicit expression (\ref{eq:Ldagger}) for $\tilde{\OP{L}}^\dagger$ that
$\int\mathrm{d}\tilde{\VEC{v}}\mathrm{d}\tilde{\VEC{\omega}}\,\tilde{\OP{L}}^\dagger G^{(0)}=0$,
so that we find $g^{(0)}$ in (\ref{eq:G0}) to be
independent of the intermediate time-scale $\vartheta$,
\begin{equation}
\label{eq:SC_order-1}
\frac{\partial g^{(0)}}{\partial\vartheta} = 0 \, .
\end{equation}
Applying the solvability condition in an analogous way
to (\ref{eq:order0}) we obtain
\begin{equation}
\label{eq:SC_order0}
\frac{\partial g^{(0)}}{\partial\tau}  + \frac{\partial g^{(1)}}{\partial\vartheta}
= \int\mathrm{d}\tilde{\VEC{v}}\mathrm{d}\tilde{\VEC{\omega}}\,\tilde{\OP{L}}^\dagger G^{(1)} \, ,
\end{equation}
where we have used
$\int\mathrm{d}\tilde{\VEC{v}}\mathrm{d}\tilde{\VEC{\omega}}\,\tilde{\OP{N}}^\dagger G^{(0)}=0$.

We can now state more precisely what it
means to derive the overdamped equation of motion
for the generating function $G_{\tilde{s}_1\tilde{s}_2\tilde{s}_3}$.
This overdamped equation should be valid only
on time-scales beyond those of $\theta$,
after the fast degrees of freedom have relaxed and have
reached their stationary distribution (on $\theta$ time-scales),
so that they can be integrated out.
The overdamped generating function is thus given by
\begin{equation}
\label{eq:limG}
g_{\tilde{s}_1\tilde{s}_2\tilde{s}_3} = \lim_{\EPS \to 0}
\int \mathrm{d}\tilde{\VEC{v}}\mathrm{d}\tilde{\VEC{\omega}} \, G_{\tilde{s}_1\tilde{s}_2\tilde{s}_3} \, ,
\end{equation}
for which we want to calculate the equation of motion
\begin{equation}
\label{eq:limdGdt}
\frac{\partial g _{\tilde{s}_1\tilde{s}_2\tilde{s}_3}}{\partial\tilde{t}} =
\lim_{\EPS \to 0}
\int \mathrm{d}\tilde{\VEC{v}}\mathrm{d}\tilde{\VEC{\omega}} \, \frac{\partial G_{\tilde{s}_1\tilde{s}_2\tilde{s}_3}}{\partial\tilde{t}} \, .
\end{equation}
The overdamped limit $\EPS=\sqrt{\tau_v/\tau_x} \to 0$
singles out the zeroth-order contribution in
$G_{\tilde{s}_1\tilde{s}_2\tilde{s}_3}$ and ${\partial G_{\tilde{s}_1\tilde{s}_2\tilde{s}_3}}/{\partial\tilde{t}}$.
It follows from (\ref{eq:expansionG}), (\ref{eq:G0}) that $g_{\tilde{s}_1\tilde{s}_2\tilde{s}_3}$
is actually identical to $g^{(0)}$.
Moreover, inserting the relations (\ref{eq:expansionG}) and (\ref{eq:Dtimes}) into (\ref{eq:limdGdt}),
and using the results we so far obtained from the perturbation expansion,
namely $\partial G^{(0)}/\partial\theta=0$,
as well as (\ref{eq:SC_order-1}) and (\ref{eq:SC_order0}),
we find
\begin{equation}
\frac{\partial g _{\tilde{s}_1\tilde{s}_2\tilde{s}_3}}{\partial\tilde{t}} =
\int\mathrm{d}\tilde{\VEC{v}}\mathrm{d}\tilde{\VEC{\omega}}\,\tilde{\OP{L}}^\dagger G^{(1)} \, .
\end{equation}
In the integral on the right-hand side the
$\partial/\partial\tilde{v}_i$ and $\partial/\partial\tilde{\omega}_i$ terms
from (\ref{eq:Ldagger}) evaluate to zero.
Our quantity of interest therefore becomes
\begin{multline}
\label{eq:dgdt}
\frac{\partial g_{\tilde{s}_1\tilde{s}_2\tilde{s}_3}}{\partial\tilde{t}} = 
-\frac{\partial J_{\tilde{v}_i}}{\partial\tilde{x}_i}
-\epsilon_{ijk} \left( \frac{\partial}{\partial\tilde{n}_i}\tilde{n}_k + \frac{\partial}{\partial\tilde{m}_i}\tilde{m}_k \right) J_{\tilde{\omega}_j}
\\ 
-\tilde{s}_1 \left[ \frac{\tilde{f}_i J_{\tilde{v}_i}}{\tilde{T}}
				  - \frac{J_{\tilde{v}_i}}{\tilde{T}}\frac{\partial\tilde{T}}{\partial\tilde{x}_i}
				  + \frac{\tilde{M}_i J_{\tilde{\omega}_i}}{\tilde{T}} \right]
\\
-\tilde{s}_3 \left( \frac{5\tilde{T}J_{\tilde{v}_i}-J_{\tilde{v}_j\tilde{v}_j\tilde{v}_i}}{2\tilde{T}^2}
				  + \frac{3\tilde{T}J_{\tilde{v}_i}-J_{\tilde{I}_{jk}\tilde{\omega}_j\tilde{\omega}_k\tilde{v}_i}}{2\tilde{T}^2} \right)
			 \frac{\partial\tilde{T}}{\partial\tilde{x}_i} \, ,
\end{multline}
with the definition
\begin{equation}
\label{eq:J}
J_{\pi(\tilde{\VEC{v}},\tilde{\VEC{\omega}})} = \int\mathrm{d}\tilde{\VEC{v}}\mathrm{d}\tilde{\VEC{\omega}}\, \pi(\tilde{\VEC{v}},\tilde{\VEC{\omega}}) G^{(1)}
\end{equation}
for integrals over (polynomial) functions $\pi(\tilde{\VEC{v}},\tilde{\VEC{\omega}})$ in
$\tilde{\VEC{v}}$ and $\tilde{\VEC{\omega}}$ multiplying $G^{(1)}$.
We remark that (\ref{eq:dgdt}) does not depend explicitly on $\tilde{s}_2$,
as $G^{(1)}$ is independent of $\tilde{s}_2$ according to (\ref{eq:order-1}),
with the consequence that in the overdamped limit the temperature variations with time
do not show up directly.
For evaluating the remaining various integrals $J_{\pi(\tilde{\VEC{v}},\tilde{\VEC{\omega}})}$
appearing in (\ref{eq:dgdt}),
it is actually not necessary to find the full solution $G^{(1)}$
of (\ref{eq:order-1}). Instead, they
can be evaluated directly from (\ref{eq:order-1})
using the known solution (\ref{eq:G0}) for $G^{(0)}$,
as shown in Appendix \ref{app:J}. The final result reads
\begin{widetext}
\begin{multline}
\label{eq:dgdtres}
\frac{\partial g_{\tilde{s}_1\tilde{s}_2\tilde{s}_3}}{\partial\tilde{t}} =
-\frac{\partial}{\partial\tilde{x}_i} 
 	\left[ (\tilde{\gamma}^{-1})_{ij} \tilde{f}_j - (\tilde{\gamma}^{-1})_{ij}\frac{\partial}{\partial\tilde{x}_j}\tilde{T} \right]
	g_{\tilde{s}_1\tilde{s}_2\tilde{s}_3}
\\
-\epsilon_{ijk} \left( \frac{\partial}{\partial\tilde{n}_i}\tilde{n}_k + \frac{\partial}{\partial\tilde{m}_i}\tilde{m}_k \right)
	\left[ (\tilde{\eta}^{-1})_{jl} \tilde{M}_l - (\tilde{\eta}^{-1})_{jl} \tilde{T} \epsilon_{plq} \left( \frac{\partial}{\partial\tilde{n}_p}\tilde{n}_q + \frac{\partial}{\partial\tilde{m}_p}\tilde{m}_q \right) \right]
	g_{\tilde{s}_1\tilde{s}_2\tilde{s}_3}
\\
-\tilde{s}_1 \left[
	\tilde{T}(\tilde{\gamma}^{-1})_{ij}\frac{\partial}{\partial\tilde{x}_j}
	\left( \frac{\tilde{f}_i}{\tilde{T}}-\frac{1}{\tilde{T}}\frac{\partial\tilde{T}}{\partial\tilde{x}_i} \right)
  + \left( (\tilde{\gamma}^{-1})_{ij}\tilde{f}_j + \tilde{T} \frac{\partial(\tilde{\gamma}^{-1})_{ij}}{\partial\tilde{x}_j} \right)
	\left( \frac{\tilde{f}_i}{\tilde{T}}-\frac{1}{\tilde{T}}\frac{\partial\tilde{T}}{\partial\tilde{x}_i} \right)
\right] g_{\tilde{s}_1\tilde{s}_2\tilde{s}_3}
\\
+ \tilde{s}_1^2 \, \tilde{T} (\tilde{\gamma}^{-1})_{ij}
	\left( \frac{\tilde{f}_i}{\tilde{T}}-\frac{1}{\tilde{T}}\frac{\partial\tilde{T}}{\partial\tilde{x}_i} \right)
	\left( \frac{\tilde{f}_j}{\tilde{T}}-\frac{1}{\tilde{T}}\frac{\partial\tilde{T}}{\partial\tilde{x}_j} \right)
	g_{\tilde{s}_1\tilde{s}_2\tilde{s}_3}
+ 2 \tilde{s}_1 \left[ \frac{\partial}{\partial\tilde{x}_i} (\tilde{\gamma}^{-1})_{ij}
	\left( \tilde{f}_j-\frac{\partial\tilde{T}}{\partial\tilde{x}_j} \right) g_{\tilde{s}_1\tilde{s}_2\tilde{s}_3} \right]
\\
- \tilde{s}_1 \left[
	\frac{(\tilde{\eta}^{-1})_{ij}}{\tilde{T}}\tilde{M}_i\tilde{M}_j + \epsilon_{ijk}(\tilde{\eta}^{-1})_{jl}
		\left( \tilde{n}_k \frac{\partial\tilde{M}_l}{\partial\tilde{n}_i} + \tilde{m}_k \frac{\partial\tilde{M}_l}{\partial\tilde{m}_i} \right)
\right] g_{\tilde{s}_1\tilde{s}_2\tilde{s}_3}
\\
+ \tilde{s}_1^2 \, \frac{(\tilde{\eta}^{-1})_{ij}}{\tilde{T}}\tilde{M}_i\tilde{M}_j g_{\tilde{s}_1\tilde{s}_2\tilde{s}_3}
+ 2 \tilde{s}_1 \left[
	\left( \frac{\partial}{\partial\tilde{n}_i}\tilde{n}_k + \frac{\partial}{\partial\tilde{m}_i}\tilde{m}_k\right)
	\epsilon_{ijk} (\tilde{\eta}^{-1})_{jl} \tilde{M}_l g_{\tilde{s}_1\tilde{s}_2\tilde{s}_3}
\right]
\\
+ \tilde{s}_3(\tilde{s}_3-1) \frac{1}{2\tilde{T}} \left(
	\frac{2U_{ik}U_{jk}}{3\tilde{\gamma}^{(k)}} + \sum_l \frac{U_{ik}U_{jk}}{\tilde{\gamma}^{(k)}+2\tilde{\gamma}^{(l)}}
  + \sum_l \frac{U_{ik}U_{jk}}{\tilde{\gamma}^{(k)}+2\tilde{\eta}^{(l)}/\tilde{I}^{(l)}}
\right)
\frac{\partial\tilde{T}}{\partial\tilde{x}_i}\frac{\partial\tilde{T}}{\partial\tilde{x}_j} g_{\tilde{s}_1\tilde{s}_2\tilde{s}_3}
\, ,
\end{multline}
\end{widetext}
where we have used on the right-hand side that
$g^{(0)}=g_{\tilde{s}_1\tilde{s}_2\tilde{s}_3}$,
and where $U_{ik}$ is defined in Eq.~(\ref{eq:diaggamma})
of Appendix \ref{app:J}.
This expression, together with its interpretation in the
following Sections constitute the main results of this paper.

\section{Overdamped dynamics}
In analogy to (\ref{eq:G000}) we obtain the overdamped limit
of the probability density
$\rho(\VEC{x},\VEC{n},\VEC{m},t)=\lim_{\EPS \to 0} \int\mathrm{d}\VEC{v}\mathrm{d}\VEC{\omega} \, p(\VEC{x},\VEC{v},\VEC{n},\VEC{m},\VEC{\omega},t)$
from $g_{000}$.
According to (\ref{eq:limG}) and (\ref{eq:limdGdt}),
we can thus read off
the overdamped equation of motion for
$\rho(\VEC{x},\VEC{n},\VEC{m},t)$
from (\ref{eq:dgdtres})
by setting $\tilde{s}_1=0$, $\tilde{s}_2=0$, $\tilde{s}_3=0$:
\begin{equation}
\label{eq:FP}
\frac{\partial\rho}{\partial t} - \OP{A}_{\mathrm{over}}^\dagger \rho = 0 \, ,
\end{equation}
with
\begin{widetext}
\begin{multline}
\label{eq:AdaggerOverdamped}
\OP{A}_{\mathrm{over}}^\dagger =
-\frac{\partial}{\partial x_i} 
 	\left[ (\gamma^{-1})_{ij} f_j - (\gamma^{-1})_{ij}\frac{\partial}{\partial x_j} k_{\mathrm{B}}T \right]
\\
-\epsilon_{ijk} \left( \frac{\partial}{\partial n_i}n_k + \frac{\partial}{\partial m_i}m_k \right)
	\left[ (\eta^{-1})_{jl} M_l - (\eta^{-1})_{jl} k_{\mathrm{B}}T \epsilon_{plq} \left( \frac{\partial}{\partial n_p}n_q + \frac{\partial}{\partial m_p}m_q \right) \right] \, ,
\end{multline}
\end{widetext}
where we switched back to dimensionful quantities.
Here and in the following, we use a sub- or superscript ``over'' to indicate the overdamped limit
and to differentiate overdamped quantities from the original ones.
The Langevin equations, which are equivalent to
the Fokker-Planck equation (\ref{eq:FP}), (\ref{eq:AdaggerOverdamped}), read
\begin{subequations}
\label{eq:langevinOverdamped}
\begin{eqnarray}
\dot{\VEC{x}} & = &
\gamma^{-1}\VEC{f} - \frac{\gamma^{-1}}{2}\frac{\partial k_{\mathrm{B}}T}{\partial\VEC{x}} + \frac{k_{\mathrm{B}}T}{2}\frac{\partial\gamma^{-1}}{\partial\VEC{x}}
\nonumber \\
&& \qquad \mbox{}+ \sqrt{2k_{\mathrm{B}}T}\gamma^{-1/2} \circ \VEC{\xi}(t) \, ,
\label{eq:xDotOverdamped} \\[1ex]
\dot{\VEC{n}} & = &
\left[ \eta^{-1} \VEC{M} + \sqrt{2k_{\mathrm{B}}T}\eta^{-1/2}\VEC{\zeta}(t) \right] \times \VEC{n} \, ,
\label{eq:nDotOverdamped} \\
\dot{\VEC{m}} & = &
\left[ \eta^{-1} \VEC{M} + \sqrt{2k_{\mathrm{B}}T}\eta^{-1/2}\VEC{\zeta}(t) \right] \times \VEC{m} \, ,
\label{eq:mDotOverdamped}
\end{eqnarray}
\end{subequations}
where $\VEC{\xi}(t)$ and $\VEC{\zeta}(t)$ are Gaussian white noise sources
(like in (\ref{eq:langevinTrans}), (\ref{eq:langevinRot}), but not related to those),
and where all the products involving noise terms, even the cross products,
are to be interpreted in the Stratonovich sense.
These Langevin equations are well-known as a model for the overdamped
translation and rotation of ellipsoidal particles \cite{mazo02}.
In particular the ``splitting'' of the translational diffusion tensor $D=k_{\mathrm{B}}T\gamma^{-1}$
into $k_{\mathrm{B}}T$ and $\gamma^{-1}$ between the two spatial derivatives in the
diffusion term of (\ref{eq:AdaggerOverdamped})
has been obtained before,
as well as the specific ``spurious drift terms''
in (\ref{eq:xDotOverdamped}) resulting from that ``splitting''
\cite{ryter81,sancho82,jayannavar95,hottovy12,yang13}.
Note that the explicit form of these ``spurious drift terms''
and the corresponding interpretation of the noise products
are a natural outcome of the systematic perturbation analysis.
They can not be obtained from the overdamped approximation,
naively disregarding inertia effects in (\ref{eq:langevinTrans}),
(\ref{eq:langevinRot}).
The overdamped approximation yields correct equations only
for homogeneous thermal environments when $T$, $\gamma$ and $\eta$
are constant.

\section{Overdamped entropy production and anomalous entropy}
Based on the overdamped Langevin equations (\ref{eq:langevinOverdamped}) we can
follow the standard reasoning of stochastic thermodynamics and
define the entropy production along stochastic trajectories as
\begin{equation}
\Delta S^{\mathrm{over}} = \Delta S_{\mathrm{p}}^{\mathrm{over}} + \Delta S_{\mathrm{env}}^{\mathrm{over}} \, .
\end{equation}
It is composed of contributions from the entropy change of the particle
$\Delta S_{\mathrm{p}}^{\mathrm{over}}$ and
from the entropy change in the environment $\Delta S_{\mathrm{env}}^{\mathrm{over}}$.
The entropy change of the particle is given as \cite{seifert05}
\begin{multline}
\label{eq:DeltaSpOverdamped}
\Delta S_{\mathrm{p}}^{\mathrm{over}} = k_{\mathrm{B}} \, \ln \rho(\VEC{x}_0,\VEC{n}_0,\VEC{m}_0,t_0)
\\
- k_{\mathrm{B}} \, \ln \rho(\VEC{x}(t),\VEC{n}(t),\VEC{m}(t),t) \, ,
\end{multline}
for a trajectory which starts at
at a point $(\VEC{x}_0,\VEC{n}_0,\VEC{m}_0)$ at time
$t_0$ and is located at a point $(\VEC{x}(t),\VEC{n}(t),\VEC{m}(t))$ at a later time $t$.
The entropy change in the environment can in principle be defined from
the heat exchanged with the environment.
However, due to the variations of temperature with position
such an identification is subtle for the translational
degrees of freedom \cite{celani12,matsuo00} and thus
the definition of entropy production in the environment
is better based on path probability ratios \cite{chetrite08}.
It reads
\begin{multline}
\label{eq:DeltaSenvOverdamped}
\Delta S_{\mathrm{env}}^{\mathrm{over}} = \int_{t_0}^{t}
\frac{1}{T}
\left[ \left( \VEC{f}-\frac{\partial k_{\mathrm{B}}T}{\partial\VEC{x}} \right) \circ\mathrm{d}\VEC{x}(t') \right.
\\
+ (\eta^{-1}\VEC{M})\cdot\VEC{M}\,\mathrm{d}t' + \sqrt{2k_{\mathrm{B}}T} \VEC{M}\eta^{-1/2}\circ\mathrm{d}\VEC{W}(t') \bigg] \, ,
\end{multline}
where $\mathrm{d}\VEC{W}$ is the increment of the Wiener process corresponding to
the Gaussian white noise $\VEC{\zeta}(t)$ in (\ref{eq:nDotOverdamped}), (\ref{eq:mDotOverdamped}).
In that way, one arrives at an entropy production which is expressed as
a sequential functional on overdamped trajectories.

We now compare the expressions (\ref{eq:DeltaSpOverdamped}) and
(\ref{eq:DeltaSenvOverdamped}), obtained from applying
stochastic thermodynamics principles ``naively'' to the
overdamped equations of motion (\ref{eq:langevinOverdamped}),
with the full result (\ref{eq:dgdtres}) from the systematic perturbation
analysis.
We first observe that the overdamped limit $\EPS \to 0$ of
the probability density $p$ appearing in
the change of particle
entropy (\ref{eq:DeltaSp}) is given by
$G^{(0)}$ evaluated at $\tilde{s}_1=0$, $\tilde{s}_2=0$, $\tilde{s}_3=0$,
which we obtain from (\ref{eq:G0}) to be
$\rho(\VEC{x},\VEC{n},\VEC{m},t)w_v(\VEC{v}|\VEC{x},t)w_\omega(\VEC{\omega}|\VEC{x},t)$
(dimensionful units).
Therefore, $\lim_{\EPS \to 0}\Delta S_{\textrm{p}}$ yields
the terms listed in (\ref{eq:DeltaSpOverdamped}), plus additional terms
involving $\log w_v w_\omega$ which
cancel precisely with the first line in (\ref{eq:DeltaSenv_split}).
We conclude that, in the overdamped limit,
the change of the particle
entropy can indeed be identified with (\ref{eq:DeltaSpOverdamped}),
while the entropy production in the environment originates from
the overdamped counterparts
$\Delta S_{\mathrm{reg}}^{\mathrm{over}}$,
$\Delta S_{\mathrm{time}}^{\mathrm{over}}$,
$\Delta S_{\mathrm{anom}}^{\mathrm{over}}$
of the entropy terms in the second line of (\ref{eq:DeltaSenv_split}).
These overdamped entropy contributions
are encoded in our main result (\ref{eq:dgdtres})
by the definition
\begin{multline}
\label{eq:gs}
g_{s_1 s_2 s_3}(\VEC{x},\VEC{n},\VEC{m},t|\VEC{x}_0,\VEC{n}_0,\VEC{m}_0,t_0)
\\
= \left\langle \exp(-s_1\Delta S_{\mathrm{reg}}^{\mathrm{over}} - s_2\Delta S_{\mathrm{time}}^{\mathrm{over}} - s_3 \Delta S_{\mathrm{anom}}^{\mathrm{over}})
\right. \\ \left.
\delta(\VEC{x}(t)-\VEC{x})\delta(\VEC{n}(t)-\VEC{n})\delta(\VEC{m}(t)-\VEC{m}) \right\rangle
\, ,
\end{multline}
analogous to (\ref{eq:Gs}).
To access and analyze
the specific form of an individual contribution we set
the $s$ variables associated with the other contributions to zero
and compare the remaining terms in (\ref{eq:dgdtres}) with the general
formulas from Appendix \ref{app:FE}.
In that way, we find the following:

The entropy production
given in (\ref{eq:DeltaSenvOverdamped})
arises precisely from the overdamped limit of the regular part
(\ref{eq:DeltaSreg}),
represented by the $\tilde{s}_1$ terms in (\ref{eq:dgdtres}).
There is no contribution from
the entropy production (\ref{eq:DeltaStime})
due time-changes of temperature,
because (\ref{eq:dgdtres}) does not explicitly depend on $\tilde{s}_2$
(as already pointed out earlier), such that
$\Delta S_{\mathrm{time}}^{\mathrm{over}}$
is bound to vanish identically.
However, the $\tilde{s}_3$ terms
yield additional contributions to the overdamped entropy production,
which are \emph{not} included in (\ref{eq:DeltaSenvOverdamped}).
From their specific functional form we infer (cf.\ Appendix \ref{app:FE})
that these contributions can not even be expressed as a sequential functional
over overdamped trajectories.
Their origin is the entropy production $\Delta S_{\mathrm{anom}}$
from (\ref{eq:DeltaSanom}) (see also (\ref{eq:Gs}))
\cite{celani12,bo14}.

Although it is not possible to explicitly write this
``anomalous entropy production'' \cite{celani12}
as an integral along paths of the overdamped dynamics,
we can still derive a number of interesting and useful results
on its \emph{average} behavior from (\ref{eq:dgdtres}),
similarly to the reasoning in the Supplementary Material of \cite{celani12}.
Setting $\tilde{s}_1=0$, $\tilde{s}_2=0$, $\tilde{s}_3=1$
(i.e.\ $s_1=0$, $s_2=0$, $s_3=1/k_{\mathrm{B}}$) we find that
$g_{001}=\langle \exp(-\Delta S_{\mathrm{anom}}^{\mathrm{over}}/k_{\mathrm{B}})\delta(\VEC{x}(t)-\VEC{x})\delta(\VEC{n}(t)-\VEC{n})\delta(\VEC{m}(t)-\VEC{m}) \rangle$
obeys the same forward equation as $g_{000}=\rho(\VEC{x},\VEC{n},\VEC{m},t)$, so that
it has the solution
\begin{multline}
\left\langle
\exp\left(-\Delta S_{\mathrm{anom}}^{\mathrm{over}}/k_{\mathrm{B}}\right) \right.
\\
\left.
\delta(\VEC{x}(t)-\VEC{x})\delta(\VEC{n}(t)-\VEC{n})\delta(\VEC{m}(t)-\VEC{m})
\right\rangle
\\
= \rho(\VEC{x},\VEC{n},\VEC{m},t) \, .
\end{multline}
Integrating over the spatial coordinates $\VEC{x}$, $\VEC{n}$, $\VEC{m}$,
we find the fluctuation relation
\begin{equation}
\left\langle
\exp\left(-\Delta S_{\mathrm{anom}}^{\mathrm{over}}/k_{\mathrm{B}}\right)
\right\rangle
= 1 \, .
\end{equation}
It follows immediately (by Jensen's inequality) that
\begin{equation}
\left\langle \Delta S_{\mathrm{anom}}^{\mathrm{over}} \right\rangle \geq 0 \, .
\end{equation}

The explicit form for the average rate of anomalous entropy
production can be obtained from (\ref{eq:dgdtres})
by observing that
$\frac{\mathrm{d}}{\mathrm{d}t}\langle \Delta S_{\mathrm{anom}}^{\mathrm{over}} \rangle
= -\frac{\mathrm{d}}{\mathrm{d}t}\int \mathrm{d}\VEC{x}\,\mathrm{d}(\VEC{n},\VEC{m}) \left. \partial g_{00s_3}/\partial s_3 \right|_{s_3=0}$
(see also the derivation of Eq.~(\ref{eq:Jav}) in Appendix \ref{app:FE}).
It reads (in dimensionful quantities)
\begin{widetext}
\begin{subequations}
\label{eq:dDeltaSanomOverdampeddt}
\begin{eqnarray}
\frac{\mathrm{d}}{\mathrm{d}t}\langle \Delta S_{\mathrm{anom}}^{\mathrm{over}} \rangle
& = & k_{\mathrm{B}} \left\langle
\frac{1}{2T} \left(
	\frac{2}{3\gamma^{(k)}} + \sum_l \frac{1}{\gamma^{(k)}+2\gamma^{(l)}}
  + \sum_l \frac{1}{\gamma^{(k)}+2\eta^{(l)}/\frac{I^{(l)}}{m}}
\right)
\left( U_{ik}\frac{\partial T}{\partial x_i} \right)
\left( U_{jk}\frac{\partial T}{\partial x_j} \right)
\right\rangle
\\
& = & k_{\mathrm{B}} \left\langle
\frac{1}{2T} \left[
\frac{2}{3}\left(\gamma^{-1}\right)_{ij}
+ \sum_l \left( \left[ \gamma+2\gamma^{(l)}\mathbb{I} \right]^{-1} \right)_{ij}
+ \sum_l \left( \left[ \gamma+\frac{2\eta^{(l)}}{I^{(l)}/m}\mathbb{I} \right]^{-1} \right)_{ij}
\right]
\frac{\partial T}{\partial x_i}\frac{\partial T}{\partial x_j}
\right\rangle
\\
& = & k_{\mathrm{B}} \int\mathrm{d}\VEC{x}\,\mathrm{d}(\VEC{n},\VEC{m})\,
\frac{\rho}{2T}
\left( \frac{\partial T}{\partial \VEC{x}} \right)\TRANS
\left[
\frac{2}{3}\gamma^{-1}
+ \sum_l \left( \gamma+2\gamma^{(l)}\mathbb{I} \right)^{-1}
+ \sum_l \left( \gamma+\frac{2\eta^{(l)}}{I^{(l)}/m}\mathbb{I} \right)^{-1}
\right]
\frac{\partial T}{\partial \VEC{x}}
\, .
\end{eqnarray}
\end{subequations}
\end{widetext}
These expressions for the average rate of anomalous entropy production
are another central result of the present paper.
We give three different, but equivalent forms.
In the first line, the entropy production is written in the coordinate frame
fixed to the particle.
Accordingly, the quantity $U_{ik}\frac{\partial T}{\partial x_i}$ is the
temperature gradient along that principal axis of the particle,
for which the friction coefficients are $\gamma^{(k)}$ and $\eta^{(k)}$
(and moment of inertia is $I^{(k)}$);
see Eq.~(\ref{eq:diaggamma}) in Appendix \ref{app:J} where
also the $U_{ik}$ are defined.
The second line represents the anomalous entropy production
in the laboratory frame of reference ($\mathbb{I}$ denotes the identity matrix).
In the third line, we switch back to vector notation
and we express the average $\langle\ldots\rangle$ over particle trajectories
explicitly as an integral over the probability density
$\rho=\rho(\VEC{x},\VEC{n},\VEC{m},t)$; note that the integral over $\mathrm{d}(\VEC{n},\VEC{m})$
is not performed independently over $\VEC{n}$ and $\VEC{m}$ but
rather represents an integral over the space of particle orientations,
parametrized by $\VEC{n}$, $\VEC{m}$.

The result (\ref{eq:dDeltaSanomOverdampeddt}) generalizes the discovery of
the anomalous entropy in \cite{celani12} in essentially two
respects. First, it covers non-trivial particle shapes,
quantifying deviations from a perfectly spherical bead by the principal
values $\gamma^{(i)}$, $\eta^{(i)}$ and $I^{(i)}/m$.
Note that the ratio $I^{(i)}/m$ does not depend on the particle
mass, as $I^{(i)}$ is proportional to $m$; it is thus a purely
geometrical factor reflecting the particle's shape.
Second, it takes into account the rotational degrees of freedom
of the Brownian particle, and reveals that rotational motion
adds to the entropy production.
Partial results about the effects of rotation on the entropic anomaly
have also been derived in \cite{lan15}.
The rotational entropy production originates from the terms
involving the rotational friction coefficients $\eta^{(i)}$, since
these terms vanish when we ``freeze'' the particle rotation
using the limit $\eta^{(i)}\to\infty$ to be left with the
translational motion only.
We can therefore identify the translational and rotational
contributions to the anomalous entropy production as
\begin{subequations}
\label{eq:kappa}
\begin{eqnarray}
\kappa_{\mathrm{trans}} & = &
\frac{k_{\mathrm{B}}\rho}{2T} \left(
	\frac{2}{3}\gamma^{-1} + \sum_l \left( \gamma+2\gamma^{(l)}\mathbb{I} \right)^{-1}
\right)
\, , \\
\kappa_{\mathrm{rot}}   & = &
\frac{k_{\mathrm{B}}\rho}{2T} \sum_l \left( \gamma+\frac{2\eta^{(l)}}{I^{(l)}/m}\mathbb{I} \right)^{-1}
\, .
\end{eqnarray}
\end{subequations}
It has been argued in \cite{celani12} that the anomalous entropy
is generated by the particle permanently transporting heat between
adjacent regions at different temperatures in the
inhomogeneous thermal environment on the fast time-scale $\tau_v$ and
associated length-scale $(k_{\mathrm{B}}T_0 m)^{1/2}/\gamma_0$ without
performing any visible displacement on the long (overdamped) time-scale
$\tau_x$. We can therefore interpret $\kappa_{\mathrm{trans}}$
and $\kappa_{\mathrm{rot}}$ as state-dependent, anisotropic heat conductivities
quantifying this process behind the anomalous entropy production \cite{lan15}.
This interpretation also explains why the rotational contribution
$\kappa_{\mathrm{rot}}$ depends on the translational friction
coefficients $\gamma$. The conducted heat is ``stored'' in
the rotational degrees of freedom, but is transported from one
temperature region to another by translation. Without translational
motion, the particle's rotation can not produce ``anomalous entropy''.

We finally remark that the rotational motion contributes
to the anomalous entropy production even if the particle is perfectly spherical.
For homogeneous spherical beads with radius $a$,
translational and rotational friction tensors as well as
the moment of inertia tensor are proportional to the unit
tensor, i.e.\ we have \cite{happel83}
$\gamma^{(1)}=\gamma^{(2)}=\gamma^{(3)}=6\pi\nu a$,
$\eta^{(1)}=\eta^{(2)}=\eta^{(3)}=8\pi\nu a^3$
and $I^{(1)}=I^{(2)}=I^{(3)}=2ma^2/5$,
so that the conductivities become isotropic,
\begin{subequations}
\label{eq:kappaSphere}
\begin{eqnarray}
\kappa_{\mathrm{trans}} & = &
\frac{5k_{\mathrm{B}}\rho}{6T\gamma^{(1)}} \,\mathbb{I}
=
\frac{5k_{\mathrm{B}}\rho}{36\pi T\nu a} \,\mathbb{I}
\, , \\
\kappa_{\mathrm{rot}}   & = &
\frac{3k_{\mathrm{B}}\rho}{2T} \frac{1}{\gamma^{(1)}+2\eta^{(1)}/\frac{I^{(1)}}{m}} \,\mathbb{I}
=
\frac{3k_{\mathrm{B}}\rho}{92\pi T\nu a} \,\mathbb{I}
\, .
\end{eqnarray}
\end{subequations}
For the translational part we just recover the
result from \cite{celani12}, while the particle rotation
gives rise to an additional contribution not described
in \cite{celani12}.
The frictional ``coefficients'' $\gamma^{(1)}$ and
$\eta^{(1)}/\frac{I^{(1)}}{m}$, representing translation and rotation,
respectively, quantify physically related effects and are
of similar magnitude. Hence,
the rotational contribution to the ``anomalous entropy'' is
well comparable to the translational part,
and is actually only by about a factor four smaller
(see also the discussion in \cite{lan15}).

\section{Prolate and oblate spheroids}
For an ellipsoidal particle
\begin{equation}
\label{eq:ellipsoid}
\frac{x_1^2}{a_1^2} + \frac{x_2^2}{a_2^2} + \frac{x_3^2}{a_3^2} = 1 
\end{equation}
with semi-axis lengths $a_1$, $a_2$, $a_3$,
the translational friction coefficients $\gamma^{(i)}$
have been calculated by Oberbeck \cite{oberbeck1876},
the rotational ones $\eta^{(i)}$ by
Edwardes \cite{edwardes1892} and Jeffery \cite{jeffery22}.
They are summarized, for instance, in \cite{brenner67}
and read
\begin{subequations}
\label{eq:gammaeta}
\begin{eqnarray}
\gamma^{(i)} & = & 16\pi\nu\frac{1}{\chi + a_i^2 \alpha_i} \quad(\mbox{for } i=1,2,3) \, ,
\\
\eta^{(1)} & = & \frac{16\pi\nu}{3}\frac{a_2^2 + a_3^2}{a_2^2\alpha_2 + a_3^2\alpha_3} \, ,
\\
\eta^{(2)} & = & \frac{16\pi\nu}{3}\frac{a_3^2 + a_1^2}{a_3^2\alpha_3 + a_1^2\alpha_1} \, ,
\\
\eta^{(3)} & = & \frac{16\pi\nu}{3}\frac{a_1^2 + a_2^2}{a_1^2\alpha_1 + a_2^2\alpha_2} \, ,
\end{eqnarray}
\end{subequations}
with
\begin{subequations}
\label{eq:chialphaDelta}
\begin{eqnarray}
\chi & = & \int_0^\infty \frac{\mathrm{d}\lambda}{\Delta(\lambda)} \, ,
\\
\alpha_i & = & \int_0^\infty \frac{\mathrm{d}\lambda}{(a_i^2 + \lambda)\Delta(\lambda)} \quad(\mbox{for } i=1,2,3) \, ,
\\
\Delta(\lambda) & = & [(a_1^2 + \lambda)(a_2^2 + \lambda)(a_3^2 + \lambda)]^{1/2} \, ,
\end{eqnarray}
\end{subequations}
and with $\nu$ being the viscosity of the medium.
The moment of inertia tensor for the ellipsoid
(\ref{eq:ellipsoid}) is
\begin{equation}
\label{eq:Im}
\frac{1}{m}\left(
\begin{array}{ccc}
I^{(1)}	& 0			& 0			\\
0		& I^{(2)}	& 0			\\
0		& 0			& I^{(3)}
\end{array}
\right)
=
\left(
\begin{array}{ccc}
\frac{a_2^2+a_3^2}{5}	& 0			 			& 0			\\
0						& \frac{a_3^2+a_1^2}{5}	& 0			\\
0						& 0						& \frac{a_1^2+a_2^2}{5}
\end{array}
\right) \, .
\end{equation}

It is clear that from these expressions we can write down
the entropic anomaly (\ref{eq:dDeltaSanomOverdampeddt})
in terms of the particle's geometry (and medium viscosity),
although the result would be quite lengthy and cumbersome and
is explicit only up to the quadratures in (\ref{eq:chialphaDelta}).
As it turns out, however, these integrals can be performed
analytically in the case of spheroids, i.e.\ ellipsoids with two equal
semi-axes, $a_1=a_2$. One distinguishes between oblate and prolate
spheroids,
\begin{eqnarray}
a_1=a_2 & > & a_3 \quad \mbox{oblate spheroid}  \, ,
\label{eq:oblate} \\
a_1=a_2 & < & a_3 \quad \mbox{prolate spheroid} \, .
\label{eq:prolate}
\end{eqnarray}

\subsection{Flat oblate spheroid}
For oblate particles, as defined in (\ref{eq:oblate}),
we obtain from (\ref{eq:chialphaDelta})
\begin{subequations}
\begin{eqnarray}
\alpha_1=\alpha_2 & = & \frac{1}{a_1^3}\left[
						\frac{-\frac{a_3}{a_1}}{1-\frac{a_3^2}{a_1^2}}
					   +\frac{\arccos\left( \frac{a_3}{a_1} \right)}{\left(1-\frac{a_3^2}{a_1^2}\right)^{3/2}}
					    \right]
\, , \\
\alpha_3 & = & \frac{2}{a_1^3}\left[
			   \frac{\frac{a_1}{a_3}}{1-\frac{a_3^2}{a_1^2}} 
			  -\frac{\arccos\left( \frac{a_3}{a_1} \right)}{\left(1-\frac{a_3^2}{a_1^2}\right)^{3/2}}
			   \right]
\, , \\
\chi & = & 	\frac{2}{a_1}\frac{\arccos\left( \frac{a_3}{a_1} \right)}{\left(1-\frac{a_3^2}{a_1^2}\right)^{1/2}}
\, .      
\end{eqnarray}
\end{subequations}
For further analysis we focus on the limiting case of
a flat oblate spheroid, assuming
\begin{equation}
\frac{a_3}{a_1} = \delta \ll 1 \, .
\end{equation}
Neglecting second and higher order
terms in $\delta$,
the translational and rotational friction coefficients
then read
\begin{subequations}
\begin{eqnarray}
\gamma^{(1)} = \gamma^{(2)} & = & \frac{32}{3} \nu a_1 \left( 1 + \frac{8}{3\pi}\delta \right)
\, , \\
\gamma^{(3)} & = & 16 \nu a_1
\, , \\
\eta^{(1)} = \eta^{(2)} & = & \frac{32}{3} \nu a_1^3
\, , \\
\eta^{(3)} & = & \frac{32}{3} \nu a_1^3 \left( 1 + \frac{4}{\pi}\delta \right)
\, .
\end{eqnarray}
\end{subequations}
The moment of inertia tensor has only zeroth
and second order terms in $\delta$,
so that to first order we find simply
\begin{equation}
\frac{1}{m}\left(
\begin{array}{ccc}
I^{(1)}	& 0			& 0			\\
0		& I^{(2)}	& 0			\\
0		& 0			& I^{(3)}
\end{array}
\right)
=
\left(
\begin{array}{ccc}
\frac{a_1^2}{5}	& 0			 		& 0			\\
0				& \frac{a_1^2}{5}	& 0			\\
0				& 0					& \frac{2a_1^2}{5}
\end{array}
\right) \, .
\end{equation}

From these expressions we can easily calculate
the heat conductivities (\ref{eq:kappa}) associated
with the anomalous entropy production to first
order in $\delta$,
\begin{subequations}
\label{eq:oblate_kappa}
\begin{align}
&
\kappa^{(1)}_{\mathrm{trans}} = \kappa^{(2)}_{\mathrm{trans}} =
\frac{k_{\mathrm{B}}\rho}{32T \nu a_1} \left( \frac{19}{8} - \frac{67}{12\pi}\delta \right)
\, , \\
&
\kappa^{(3)}_{\mathrm{trans}} =
\frac{k_{\mathrm{B}}\rho}{32T \nu a_1} \left( \frac{13}{7} - \frac{64}{49\pi}\delta \right)
\, , \\
&
\kappa^{(1)}_{\mathrm{rot}} = \kappa^{(2)}_{\mathrm{rot}} =
\frac{k_{\mathrm{B}}\rho}{32T \nu a_1} \left( \frac{23}{44} - \frac{2201}{2178\pi}\delta \right)
\, , \\
&
\kappa^{(3)}_{\mathrm{rot}} =
\frac{k_{\mathrm{B}}\rho}{32T \nu a_1} \left( \frac{147}{299} - \frac{120}{169\pi}\delta \right)
\, ,
\end{align}
\end{subequations}
where the $\kappa^{(k)}_{\mathrm{trans}}$ $(k=1,2,3)$
are the translational heat conductivities
along the principal axes of the particle defined via
$(\kappa_{\mathrm{trans}})_{ij} = U_{ik}U_{jk}\kappa^{(k)}_{\mathrm{trans}}$,
and likewise for the rotational conductivities $\kappa^{(k)}_{\mathrm{rot}}$.
We find that both, translational and rotational contributions to the
anomalous entropy production are comparable in magnitude, with the translational
one being about four times larger than the rotational one.

\subsection{Thin prolate spheroid}
For the prolate particles from (\ref{eq:prolate}), we
can write (\ref{eq:chialphaDelta}) as
\begin{subequations}
\begin{eqnarray}
\alpha_1=\alpha_2 & = & \frac{1}{a_3^3}\left[
						\frac{\frac{a_3^2}{a_1^2}}{1-\frac{a_1^2}{a_3^3}}
					   -\frac{\arccosh\left( \frac{a_3}{a_1} \right)}{\left(1-\frac{a_1^2}{a_3^2}\right)^{3/2}}
					    \right]
\, , \\
\alpha_3 & = & \frac{2}{a_3^3}\left[
			   \frac{-1}{1-\frac{a_1^2}{a_3^2}} 
			  +\frac{\arccosh\left( \frac{a_3}{a_1} \right)}{\left(1-\frac{a_1^2}{a_3^2}\right)^{3/2}}
			   \right]
\, , \\
\chi & = & 	\frac{2}{a_3}\frac{\arccosh\left( \frac{a_3}{a_1} \right)}{\left(1-\frac{a_1^2}{a_3^2}\right)^{1/2}}
\, .      
\end{eqnarray}
\end{subequations}
Considering thin prolate spheroids with
\begin{equation}
\frac{a_1}{a_3} = \delta \ll 1 \, ,
\end{equation}
we obtain to lowest order in $\delta$
\begin{subequations}
\label{eq:prolate_gammaeta}
\begin{eqnarray}
\gamma^{(1)} = \gamma^{(2)} & = & \frac{16\pi\nu a_3}{2\ln 2+1-2\ln\delta}
\, , \\
\gamma^{(3)} & = &  \frac{8\pi\nu a_3}{2\ln 2-1-2\ln\delta}
\, , \\
\eta^{(1)} = \eta^{(2)} & = & \frac{16\pi\nu a_3^3}{3(2\ln 2-1-2\ln\delta)}
\, , \\
\eta^{(3)} & = & \frac{16}{3}\pi\nu a_3^3 \delta^2
\, .
\end{eqnarray}
\end{subequations}
These friction coefficients asymptotically vanish
in the limit $\delta \to 0$ (with a logarithmic approach to 0, except for $\eta^{(3)}$),
because then the prolate particle more and more
resembles a one-dimensional rod-like object which
experiences practically no friction when moving through
the fluid. It is easy to verify though that even
for $\delta \to 0$ inertia effects remain negligible
compared to viscous friction forces, since
the particle mass, being proportional to the
particle volume, decreases much faster with $\delta \to 0$
than the friction coefficients (\ref{eq:prolate_gammaeta}).
In other words, the condition $\tau_v/\tau_x \ll 1$
[see (\ref{eq:eps})]
as a prerequisite for the overdamped limit is fulfilled
for arbitrarily small $\delta$.
The moment of inertia tensor $I/m$ as well
has entries which vanish asymptotically as $\delta \to 0$
so that we here have to 
keep the second-order terms,
\begin{equation}
\label{eq:prolate_Im}
\frac{1}{m}\left(
\begin{array}{ccc}
I^{(1)}	& 0			& 0			\\
0		& I^{(2)}	& 0			\\
0		& 0			& I^{(3)}
\end{array}
\right)
=
\left(
\begin{array}{ccc}
\frac{a_3^2 (1+\delta^2)}{5}	& 0			 					& 0			\\
0								& \frac{a_3^2 (1+\delta^2)}{5}	& 0			\\
0								& 0								& \frac{2a_3^2}{5} \delta^2
\end{array}
\right) \, .
\end{equation}

Calculating the heat conductivities
$(\kappa_{\mathrm{trans}})_{ij} = U_{ik}U_{jk}\kappa^{(k)}_{\mathrm{trans}}$ and
$(\kappa_{\mathrm{rot}})_{ij} = U_{ik}U_{jk}\kappa^{(k)}_{\mathrm{rot}}$
from (\ref{eq:kappa}) using the expansions
(\ref{eq:prolate_gammaeta}), (\ref{eq:prolate_Im}),
we find that the leading order terms diverge with $\ln (1/\delta)$.
Explicitly, the results read
\begin{subequations}
\begin{align}
&
\kappa^{(1)}_{\mathrm{trans}} = \kappa^{(2)}_{\mathrm{trans}} =
\frac{k_{\mathrm{B}}\rho}{32T \nu a_3} \left( \frac{11}{3\pi}\ln\frac{1}{\delta} + \frac{11\ln 2 + 4}{3\pi} \right)
\, , \\
&
\kappa^{(3)}_{\mathrm{trans}} =
\frac{k_{\mathrm{B}}\rho}{32T \nu a_3} \left( \frac{28}{5\pi}\ln\frac{1}{\delta} + \frac{2(70\ln 2 - 19)}{25\pi} \right)
\, , \\
&
\kappa^{(1)}_{\mathrm{rot}} = \kappa^{(2)}_{\mathrm{rot}} =
\frac{k_{\mathrm{B}}\rho}{32T \nu a_3} \left( \frac{12}{13\pi}\ln\frac{1}{\delta} + \frac{3(260\ln 2 + 99)}{845\pi} \right)
\, , \\
&
\kappa^{(3)}_{\mathrm{rot}} =
\frac{k_{\mathrm{B}}\rho}{32T \nu a_3} \left( \frac{24}{23\pi}\ln\frac{1}{\delta} + \frac{3(40\ln 2+3)}{115\pi} \right)
\, ,
\end{align}
\end{subequations}

Comparing with the corresponding results (\ref{eq:oblate_kappa})
for the flat oblate spheroid, we see that the anomalous
entropy production rate typically is
larger for the thin prolate particle,
due to the logarithmic divergence of the conductivities.

\section{Slightly deformed sphere}
Another interesting case to consider is an ellipsoid with
almost identical semi-axes, i.e.\ a particle
slightly deformed from perfect spherical shape.
We fix $a_1$ and set
\begin{equation}
\label{eq:delta2delta3}
a_2 = a_1 ( 1 + \delta_2 )
\, , \quad
a_3 = a_1 ( 1 + \delta_3 )
\, ,
\end{equation}
assuming
\begin{equation}
\delta_2 \ll 1
\, , \quad
\delta_3 \ll 1
\, .
\end{equation}
Plugging (\ref{eq:delta2delta3}) into (\ref{eq:chialphaDelta}),
we can now perform the integrations
by expanding the integrands in $\delta_2$, $\delta_3$, and
calculate the friction coefficients from (\ref{eq:gammaeta})
to a desired order in $\delta_2$, $\delta_3$.
The first-order results read
\begin{subequations}
\begin{eqnarray}
\gamma^{(1)} & = & 6\pi\nu a_1 \left( 1 + \frac{2}{5}\delta_2 + \frac{2}{5}\delta_3 \right)
\, , \\
\gamma^{(2)} & = & 6\pi\nu a_1 \left( 1 + \frac{1}{5}\delta_2 + \frac{2}{5}\delta_3 \right)
\, , \\
\gamma^{(3)} & = & 6\pi\nu a_1 \left( 1 + \frac{2}{5}\delta_2 + \frac{1}{5}\delta_3 \right)
\, , \\
\eta^{(1)} & = & 8\pi\nu a_1^3 \left( 1 + \frac{6}{5}\delta_2 + \frac{6}{5}\delta_3 \right)
\, , \\
\eta^{(2)} & = & 8\pi\nu a_1^3 \left( 1 + \frac{3}{5}\delta_2 + \frac{6}{5}\delta_3 \right)
\, , \\
\eta^{(3)} & = & 8\pi\nu a_1^3 \left( 1 + \frac{6}{5}\delta_2 + \frac{3}{5}\delta_3 \right)
\, .
\end{eqnarray}
\end{subequations}
Obviously, the well-known isotropic Stokes friction coefficients for translation
and rotation \cite{happel83} of a perfectly spherical
particle are recovered in the limit $\delta_2 \to 0$, $\delta_3 \to 0$.

For the anomalous heat conductivities $\kappa_{\mathrm{trans}}$,
$\kappa_{\mathrm{rot}}$ [see (\ref{eq:kappa})] we further need
the moment of inertia tensor $I/m$, which is 
easily obtained by inserting (\ref{eq:delta2delta3}) into (\ref{eq:Im}).
We finally find, again to first order in $\delta_2$ and $\delta_3$,
\begin{subequations}
\begin{eqnarray}
\kappa^{(1)}_{\mathrm{trans}} & = & \frac{5k_{\mathrm{B}}\rho}{36\pi T\nu a_1} \left( 1 - \frac{28}{75}\delta_2 - \frac{28}{75}\delta_3 \right)
\, , \\
\kappa^{(2)}_{\mathrm{trans}} & = & \frac{5k_{\mathrm{B}}\rho}{36\pi T\nu a_1} \left( 1 - \frac{19}{75}\delta_2 - \frac{28}{75}\delta_3 \right)
\, , \\
\kappa^{(3)}_{\mathrm{trans}} & = & \frac{5k_{\mathrm{B}}\rho}{36\pi T\nu a_1} \left( 1 - \frac{28}{75}\delta_2 - \frac{19}{75}\delta_3 \right)
\, , \\
\kappa^{(1)}_{\mathrm{rot}} & = & \frac{3k_{\mathrm{B}}\rho}{92\pi T\nu a_1} \left( 1 - \frac{118}{345}\delta_2 - \frac{118}{345}\delta_3 \right)
\, , \\
\kappa^{(2)}_{\mathrm{rot}} & = & \frac{3k_{\mathrm{B}}\rho}{92\pi T\nu a_1} \left( 1 - \frac{109}{345}\delta_2 - \frac{118}{345}\delta_3 \right)
\, , \\
\kappa^{(3)}_{\mathrm{rot}} & = & \frac{3k_{\mathrm{B}}\rho}{92\pi T\nu a_1} \left( 1 - \frac{118}{345}\delta_2 - \frac{109}{345}\delta_3 \right)
\, .
\end{eqnarray}
\end{subequations}
For $\delta_2=0$, $\delta_3=0$, we recover the results
for a spherical bead, already calculated in (\ref{eq:kappaSphere}).
Note that our observation from the spherical case, that translational
and rotational contributions to ``anomalous'' entropy
production are well comparable (about a factor four difference),
also applies for near-spherical particles.
Deviations from perfect spherical shape lead to similar corrections
for both, translational and rotational ``anomalous'' entropy.

\section{Conclusions}
In the present paper we
analyze the thermodynamic properties of a single Brownian particle
of non-spherical shape, for which rotational degrees of freedom
play a non-negligible role.
The main aim is to understand how the stochastic thermodynamics
is affected by ``coarse-graining'' the level of description
of the particle motion when performing the overdamped limit
to ``integrate out'' the fast velocity degrees of freedom.
This question is of particular interest in case that the
surrounding heat bath is heterogeneous, a situation which
is known to be non-trivial already for the particle's equations of motion
\cite{ryter81,sancho82,jayannavar95,hottovy12,yang13,widder89,freidlin04,lau07}.
A central quantity for such an analysis is the
trajectory-wise entropy production of the particle
defined according to stochastic thermodynamics \cite{seifert12}.
For translational Brownian motion,
it has been discovered in \cite{celani12} that
the overdamped limit of this entropy production
generates an ``anomalous'' contribution, which is not
captured by the statistics of the overdamped trajectories. 

Here, we analyze in detail the effects of a non-spherical particle
shape and of the Brownian rotation of such particles.
Starting from the standard entropy production of stochastic thermodynamics
(extended to include rotational degrees of freedom, see Sec.~II)
on the level of the full-fledged description of the particle
dynamics, we perform the overdamped limit using singular perturbation theory
(Sec.~III). As our main result we find that the rotational Brownian motion
not only yields a ``standard'' contribution to entropy production which
is consistent with the overdamped approximation
(where one simply disregards velocity degrees of freedom),
but in addition also generates an ``anomalous'' entropy
which can not be expressed as a functional along overdamped trajectories
[see Eqs.~(\ref{eq:dDeltaSanomOverdampeddt}) and (\ref{eq:kappa})].
This ``anomalous'' contribution to entropy production
from the particle's rotation is comparable in magnitude
to the ``anomalous'' entropy generated by translational motion.

We remark that our starting equations (\ref{eq:langevinTrans}),
(\ref{eq:langevinRot}) to model the particle's Brownian motion
do not contain any hydrodynamic coupling between translational
and rotational degrees of freedom.
Such couplings would be relevant,
for instance, for particles with a helical shape. For that reason
our analysis is restricted to the class of ellipsoidal
(and other rod- and disk-like) Brownian particles.
We expect that the more general case of hydrodynamic couplings
between translation and rotation will also induce couplings
between these degrees of freedom in the entropy production
and thus lead to an additional ``anomalous'' contribution.
The details, however, remain to be revealed in future work.

Potential applications of the present findings include
the influence of rotational Brownian motion and the associated
``anomalous'' entropy production in inhomogeneous thermal
environments on optimal time-dependent
protocols, realized by external forces 
to optimize a specific quantity
of interest during a finite-time process
\cite{bo13}, on the efficiency of microscopic stochastic heat
engines \cite{bo13a,blickle12}, and even on its
universal fluctuations discovered
recently in \cite{verley14}.

\begin{acknowledgments}
The authors would like to thank Stefano Bo,
Antonio Celani and Yueheng Lan
for valuable discussions. Financial support
by the Swedish Science Council under the grants
621-2012-2982 and 621-2013-3956 is acknowledged.
\end{acknowledgments}

\appendix

\section{Entropy production}
\label{app:EP}
In Sec.~\ref{sec:dynamicsEP} of the main text,
we derive the entropy production in the environment
(\ref{eq:DeltaSenv}) from the heat exchanged between particle and
thermal environment. In this Appendix, we summarize the relevant results
which show that (\ref{eq:DeltaSenv}) can also be obtained from the
ratio between the probability for observing a certain particle trajectory
and the probability for observing its time-reversed counterpart \cite{chetrite08}.

We consider a set of stochastic equations of the form
\begin{subequations}
\label{eq:langevinqp}
\begin{eqnarray}
\mathrm{d}q_i & = & \mu_{ij} p_j \mathrm{d}t
\, , \\
\mathrm{d}p_i & = & \left( -\Gamma_{ij} p_j + \bar{u}_i \right)\mathrm{d}t + \beta_{ij}\circ\mathrm{d}W_j
\, ,
\end{eqnarray}
\end{subequations}
where $q_i$ are space-like coordinates, transforming as
$q_i \to q_i$ if time is reversed, and $p_i$
are velocity-like coordinates with $p_i \to -p_i$
under time-reversal.
The dynamics of the space-like coordinates is linear
in the $p_i$ with a tensor $\mu_{ij}$ which may depend on
$q_i$ only (not on $p_i$).
The deterministic part of the dynamics for the velocity-like coordinates $p_i$
has a dissipative contribution $-\Gamma_{ij}p_j$ and
a part $\bar{u}_i$ collecting external forces, which
is assumed to transform as $\bar{u}_i \to \bar{u}_i$
under time-reversal.
The tensor $\beta_{ij}$ defines the strength of the thermal noise, where
the stochastic noise itself is represented by the increments of mutually
independent Wiener processes $\mathrm{d}W_i$;
the products between $\beta_{ij}$ and $\mathrm{d}W_j$
are to be interpreted in Stratonovich sense.
The diffusion tensor $D_{ij}$ resulting from these
noise terms is given by $D_{ij}=\beta_{ik}\beta_{jk}$,
where we assume that its inverse $(D^{-1})_{ij}$ exists.
Note that the structure of
the stochastic differential equations (\ref{eq:langevinqp}) covers the
Langevin equations (\ref{eq:langevinTrans}) and
(\ref{eq:langevinRot}) for translational and rotational
Brownian motion.

It can be shown
(see, e.g., \cite{chetrite08} or the Supplementary Material of \cite{celani12})
that the probability
$P[\VEC{q}(t),\VEC{p}(t)|\VEC{q}_0,\VEC{p}_0]$ for a specific trajectory
$(\VEC{q}(t),\VEC{p}(t))$
starting at $(\VEC{q}_0,\VEC{p}_0)$ at time $t_0$ and ending at
the point $(\VEC{q}_1,\VEC{p}_1)$ at a later time $t_1$
is related to the probability
$\hat{P}[\hat{\VEC{q}}(t),\hat{\VEC{p}}(t)|\hat{\VEC{q}}_0,\hat{\VEC{p}}_0]$
for the time-reversed trajectory [i.e.\ with
$(\hat{\VEC{q}}_0,\hat{\VEC{p}}_0)=(\VEC{q}_1,-\VEC{p}_1)$
and $(\hat{\VEC{q}}_1,\hat{\VEC{p}}_1)=(\VEC{q}_0,-\VEC{p}_0)$]
according to
\begin{multline}
\label{eq:PtoP}
\frac{\hat{P}[\hat{\VEC{q}}(t),\hat{\VEC{p}}(t)|\hat{\VEC{q}}_0,\hat{\VEC{p}}_0]}
	 {P[\VEC{q}(t),\VEC{p}(t)|\VEC{q}_0,\VEC{p}_0]}
\\
= \exp\left\{
- \int_{t_0}^{t_1} \left[ 2 (D^{-1})_{ij}\Gamma_{ik}p_k\left( \bar{u}_j-\dot{p}_j \right)
				 			- \frac{\partial \bar{u}_i}{\partial p_i} \right] \mathrm{d}t
\right\}
\, .
\end{multline}

We observe that for subsets of coordinates with noise sources
which are statistically independent, the inverse diffusion
tensor $D^{-1}$ takes block-diagonal structure, so
that the contributions of these subsets in the
exponent are additive, even though their deterministic forces
may depend on the whole set of coordinates.
As the Gaussian noise sources in (\ref{eq:langevinTrans})
and (\ref{eq:langevinRot}) are independent, we can
therefore focus on translation and rotation separately
to determine their contributions to the exponent in (\ref{eq:PtoP}).

For the translational motion (\ref{eq:langevinTrans}) we
identify $\VEC{q}=\VEC{x}$ and $\VEC{p}=\VEC{v}$ and find by comparison with
(\ref{eq:langevinqp})
\begin{subequations}
\begin{eqnarray}
\mu_{ij} & =& \delta_{ij}
\, , \quad
\Gamma_{ij} = \frac{\gamma_{ij}}{m}
\, , \\
\bar{u}_i & = & \frac{1}{m}f_i
\, , \\
D_{ij} & = & \frac{2k_{\mathrm{B}}T}{m^2}\gamma_{ij} \, .
\end{eqnarray}
\end{subequations}
The exponent in (\ref{eq:PtoP}) thus reads
$2\left(D^{-1}\right)_{ij}\Gamma_{ik}p_k\left( \bar{u}_j - \dot{p}_j \right) - \partial \bar{u}_i/\partial p_i
= (v_i f_i - m v_i\dot{v}_i)/(k_{\mathrm{B}}T)$ and
corresponds exactly to the translational contribution in (\ref{eq:DeltaSenv}),
up to a factor $1/k_{\mathrm{B}}$.

Likewise, the rotational motion (\ref{eq:langevinRot})
matches the equations (\ref{eq:langevinqp}) for
$\VEC{q}=(\VEC{n},\VEC{m})$, $\VEC{p}=\VEC{\omega}$ and
\begin{subequations}
\begin{eqnarray}
\mu_{ij} & = &
\left(
\begin{array}{cc}
\epsilon_{ijk}n_k	& 0					\\
0					& \epsilon_{ijk}m_k
\end{array}
\right)
\, , \\
\Gamma_{ij} & = & (I^{-1})_{ik}\eta_{kj}
\, , \\
\bar{u}_i & = & (I^{-1})_{ij}M_j - (I^{-1})_{ij}\epsilon_{jkl}\omega_k I_{lm} \omega_m
\, , \\
D_{ij} & = & 2k_{\mathrm{B}}T (I^{-1})_{ik}(I^{-1})_{lj}\eta_{kl} \, .
\end{eqnarray}
\end{subequations}
We can then calculate its contribution to the exponent to read
$2\left(D^{-1}\right)_{ij}\Gamma_{ik}p_k\left( \bar{u}_j - \dot{p}_j \right) - \partial \bar{u}_i/\partial p_i
=(\omega_i M_i-I_{ij}\omega_i\dot{\omega}_j)/(k_{\mathrm{B}}T)$,
which is equivalent to the rotational part in (\ref{eq:DeltaSenv}),
again up to a factor $1/k_{\mathrm{B}}$.

Summarizing, we therefore arrive at the central result of this Appendix,
\begin{equation}
\frac{\hat{P}}{P} = e^{-\Delta S_{\mathrm{env}}/k_{\mathrm{B}}}
\, ,
\end{equation}
relating the entropy production in the environment $\Delta S_{\mathrm{env}}$
as defined in (\ref{eq:DeltaSenv}) to the ratio of forward and backward
path probabilities.
Finally, from here it is straightforward to verify \cite{seifert05}
that for the total entropy production from (\ref{eq:DeltaS})
the integral fluctuation relation
\begin{equation}
\left\langle e^{-\Delta S/k_{\mathrm{B}}} \right\rangle = 1
\, ,
\end{equation}
is fulfilled, as already stated in
(\ref{eq:FT}).
The average $\langle\ldots\rangle$
is taken over all trajectories starting
from a fixed initial condition.

\section{Evaluation of the $G^{(1)}$ integrals $J_{\pi(\tilde{\VEC{v}},\tilde{\VEC{\omega}})}$}
\label{app:J}
We here describe how to evaluate the integrals
\begin{subequations}
\label{eq:Jlist}
\begin{eqnarray}
J_{\tilde{v}_i} & = &
\int\mathrm{d}\tilde{\VEC{v}}\mathrm{d}\tilde{\VEC{\omega}}\, \tilde{v}_i G^{(1)} \, ,
\label{eq:Jv} \\
J_{\tilde{\omega}_i} & = &
\int\mathrm{d}\tilde{\VEC{v}}\mathrm{d}\tilde{\VEC{\omega}}\, \tilde{\omega}_i G^{(1)} \, ,
\label{eq:Jomega} \\
J_{\tilde{v}_j\tilde{v}_j\tilde{v}_i} & = &
\int\mathrm{d}\tilde{\VEC{v}}\mathrm{d}\tilde{\VEC{\omega}}\, \tilde{v}_j\tilde{v}_j\tilde{v}_i G^{(1)} \, ,
\label{eq:Jvvv} \\
J_{\tilde{I}_{jk}\tilde{\omega}_j\tilde{\omega}_k\tilde{v}_i} & = &
\int\mathrm{d}\tilde{\VEC{v}}\mathrm{d}\tilde{\VEC{\omega}}\, \tilde{I}_{jk}\tilde{\omega}_j\tilde{\omega}_k\tilde{v}_i G^{(1)} \, ,
\label{eq:JIomegaomegav}
\end{eqnarray}
\end{subequations}
appearing in (\ref{eq:dgdt}),
without calculating $G^{(1)}$ explicitly.
The basic idea is to multiply the order $\EPS^{-1}$ equation
(\ref{eq:order-1}) by a polynomial $\Pi(\tilde{\VEC{v}},\tilde{\VEC{\omega}})$,
integrate over $\tilde{\VEC{v}}$, $\tilde{\VEC{\omega}}$, and
rewrite the left-hand side in terms of the operator $\tilde{\OP{M}}$
adjoint to $\tilde{\OP{M}}^\dagger$ (see (\ref{eq:Mdagger})).
In that way we obtain equations of the form
\begin{equation}
\label{eq:calcJs}
\int \mathrm{d}\tilde{\VEC{v}}\mathrm{d}\tilde{\VEC{\omega}} \, (\tilde{\OP{M}}\Pi)G^{(1)}
= -\int \mathrm{d}\tilde{\VEC{v}}\mathrm{d}\tilde{\VEC{\omega}} \, \Pi(\tilde{\OP{L}}^\dagger G^{(0)}) \, ,
\end{equation}
where we remembered that $\partial G^{(0)}/\partial\vartheta=0$ according to (\ref{eq:SC_order-1}).
For suitable choice of $\Pi$, the operation $\tilde{\OP{M}}\Pi$
in the left-hand integral
may then reproduce one of the
desired polynomials from (\ref{eq:Jlist})
(or a linear combination of several such polynomials),
while the right-hand side can be calculated straightforwardly using
the explicit expression (\ref{eq:Ldagger}) for $\tilde{\OP{L}}^\dagger$
and the known solution (\ref{eq:G0}) for $G^{(0)}$.

For instance, taking $\Pi=\tilde{v}_i$ we get
$\tilde{\OP{M}}\Pi=-\tilde{\gamma}_{ij}\tilde{v}_j$, such that after
evaluation of the corresponding right-hand side integral
$-\int \mathrm{d}\tilde{\VEC{v}}\mathrm{d}\tilde{\VEC{\omega}} \, \tilde{v}_i(\tilde{\OP{L}}^\dagger G^{(0)})$
in (\ref{eq:calcJs}) we find
\begin{equation}
\label{eq:Jvres}
J_{\tilde{v}_i} =
-(\tilde{\gamma}^{-1})_{ij}\tilde{T}\frac{\partial g^{(0)}}{\partial\tilde{x}_j}
-(\tilde{s}_1-1)(\tilde{\gamma}^{-1})_{ij} \left( \tilde{f}_j - \frac{\partial\tilde{T}}{\partial\tilde{x}_j} \right) g^{(0)} \, .
\end{equation}
Similarly, with the choice $\Pi=\tilde{\omega}_i$ we obtain
\begin{multline}
\label{eq:Jomegares}
J_{\tilde{\omega}_i} =
-(\tilde{\eta}^{-1})_{ik}\tilde{T}\epsilon_{jkl}
	\left( \frac{\partial}{\partial\tilde{n}_j}\tilde{n}_l + \frac{\partial}{\partial\tilde{m}_j}\tilde{m}_l \right) g^{(0)}
\\
-(\tilde{s}_1-1)(\tilde{\eta}^{-1})_{ij}\tilde{M}_j g^{(0)} \, .
\end{multline}

The two remaining integrals (\ref{eq:Jvvv}) and (\ref{eq:JIomegaomegav}) are
more cumbersome to calculate because of the third-order polynomials involved.
For $J_{\tilde{v}_j\tilde{v}_j\tilde{v}_i}$ it turns out that
we have to choose $\Pi=A_{ijkl}\tilde{v}_j\tilde{v}_k\tilde{v}_l$,
where the tensor $A_{ijkl}$ has to be determined such that in 
$\tilde{\OP{M}}\Pi=
	-A_{ijkl}( \tilde{\gamma}_{jm}\tilde{v}_m\tilde{v}_k\tilde{v}_l
	  +\tilde{\gamma}_{km}\tilde{v}_m\tilde{v}_j\tilde{v}_l
	  +\tilde{\gamma}_{lm}\tilde{v}_m\tilde{v}_j\tilde{v}_k )
+ 2\tilde{T}A_{ijkl}(\tilde{\gamma}_{jk}\tilde{v}_l+\tilde{\gamma}_{jl}\tilde{v}_k+\tilde{\gamma}_{kl}\tilde{v}_j)$
the sum over polynomials of third degree in $\tilde{\VEC{v}}$-components reduces to
$\tilde{v}_j\tilde{v}_j\tilde{v}_i$.
To construct the explicit form of $A_{ijkl}$, it is convenient
to diagonalize $\tilde{\gamma}$ in $\tilde{\OP{M}}$ by
\begin{equation}
\label{eq:diaggamma}
(U\TRANS\tilde{\gamma}U)_{ij} = \tilde{\gamma}^{(i)}\delta_{ij} \, ,
\end{equation}
where $U$ is a symmetric tensor and
$\tilde{\gamma}^{(i)}$ are the eigenvalues of $\tilde{\gamma}$.
It is then straightforward to identify
\begin{equation}
\label{eq:A}
A_{ijkl} = - \frac{U_{im}U_{jm}U_{kn}U_{ln}}{\tilde{\gamma}^{(m)}+2\tilde{\gamma}^{(n)}} \, .
\end{equation}
Note that the sum here is over $m$ and $n$, and that $A_{ijkl}$
obeys the symmetries $A_{ijkl}=A_{jikl}=A_{ijlk}$.
With this expression for $A_{ijkl}$ we can now calculate the two integrals
$\int \mathrm{d}\tilde{\VEC{v}}\mathrm{d}\tilde{\VEC{\omega}} \, (\tilde{\OP{M}}A_{ijkl}\tilde{v}_j\tilde{v}_k\tilde{v}_l)G^{(1)}$
and
$-\int \mathrm{d}\tilde{\VEC{v}}\mathrm{d}\tilde{\VEC{\omega}} \, A_{ijkl}\tilde{v}_j\tilde{v}_k\tilde{v}_l(\tilde{\OP{L}}^\dagger G^{(0)})$
to find
\begin{multline}
\label{eq:Jvvvres}
J_{\tilde{v}_j\tilde{v}_j\tilde{v}_i} =
5\tilde{T}J_{\tilde{v}_i} + (\tilde{s}_3-1)\tilde{T}
	\left( \frac{2U_{ik}U_{jk}}{3\tilde{\gamma}^{(k)}} \right.
	\\
	\left. + \sum_l \frac{U_{ik}U_{jk}}{\tilde{\gamma}^{(k)}+2\tilde{\gamma}^{(l)}} \right)
	\frac{\partial\tilde{T}}{\partial\tilde{x}_j} g^{(0)}\, .
\end{multline}
The sum over $l$ is specified explicitly, since this index appears
only once; apart from that summation over double indices is still understood.
To arrive at the simple form (\ref{eq:Jvvvres}) we made use of (\ref{eq:Jvres})
and of the relations
$2A_{ilkl}+A_{ikll}=-\frac{2U_{ik}U_{jk}}{3\tilde{\gamma}^{(k)}}
-\sum_l\frac{U_{ik}U_{jk}}{\tilde{\gamma}^{(k)}+2\tilde{\gamma}^{(l)}}$
and
$(4A_{iklj}+2A_{ijkl})\tilde{\gamma}_{kl}+(2A_{ilkl}+A_{ikll})\tilde{\gamma}_{kj}=-5\delta_{ij}$,
which can be proven by using (\ref{eq:diaggamma}), (\ref{eq:A}).

Finally, the calculation of $J_{\tilde{I}_{jk}\tilde{\omega}_j\tilde{\omega}_k\tilde{v}_i}$
from (\ref{eq:JIomegaomegav})
proceeds completely analogously. The proper choice for $\Pi$ turns out to be
$B_{ijkl}\tilde{v}_j\tilde{\omega}_k\tilde{\omega}_l$, with
\begin{equation}
\label{eq:B}
B_{ijkl} = - \frac{U_{im}U_{jm}V_{kn}V_{ln}\tilde{I}^{(n)}}{\tilde{\gamma}^{(m)}+2\tilde{\eta}^{(n)}/\tilde{I}^{(n)}}
\end{equation}
to guarantee that $\tilde{\OP{M}}B_{ijkl}\tilde{v}_j\tilde{\omega}_k\tilde{\omega}_l = \tilde{I}_{jk}\tilde{\omega}_j\tilde{\omega}_k\tilde{v}_i
+ (\mbox{terms linear in $\tilde{\VEC{v}}$-components})$.
In determining (\ref{eq:B}) we have assumed that $\tilde{I}$ and $\tilde{\eta}$ are
diagonalized simultaneously by $V$
\cite{Note3},
\begin{equation}
(V\TRANS\tilde{I}V)_{ij} = \tilde{I}^{(i)}\delta_{ij} \, ,
\quad
(V\TRANS\tilde{\eta}V)_{ij} = \tilde{\eta}^{(i)}\delta_{ij} \, .
\end{equation}
Then, evaluating
$\int \mathrm{d}\tilde{\VEC{v}}\mathrm{d}\tilde{\VEC{\omega}} \, (\tilde{\OP{M}}B_{ijkl}\tilde{v}_j\tilde{\omega}_k\tilde{\omega}_l)G^{(1)}$
and
$-\int \mathrm{d}\tilde{\VEC{v}}\mathrm{d}\tilde{\VEC{\omega}} \, B_{ijkl}\tilde{v}_j\tilde{\omega}_k\tilde{\omega}_l(\tilde{\OP{L}}^\dagger G^{(0)})$
we obtain
\begin{multline}
\label{eq:JIomegaomegavres}
J_{\tilde{I}_{jk}\tilde{\omega}_j\tilde{\omega}_k\tilde{v}_i} =
3\tilde{T}J_{\tilde{v}_i}
\\
+ (\tilde{s}_3-1)\tilde{T}
\sum_l \frac{U_{ik}U_{jk}}{\tilde{\gamma}^{(k)}+2\tilde{\eta}^{(l)}/\tilde{I}^{(l)}}\frac{\partial\tilde{T}}{\partial\tilde{x}_j} g^{(0)} \, .
\end{multline}
Again, we have simplified (\ref{eq:JIomegaomegavres})
by using (\ref{eq:Jvres}), and by observing that
$B_{ijkl}(\tilde{I}^{-1})_{kl}=-\sum_l \frac{U_{im}U_{jm}}{\tilde{\gamma}^{(m)}+2\tilde{\eta}^{(l)}/\tilde{I}^{(l)}}$
and
$2B_{ijkl}(\tilde{I}^{-1})_{km}(\tilde{I}^{-1})_{ln}\tilde{\eta}_{mn}+B_{iklm}(\tilde{I}^{-1})_{lm}\tilde{\gamma}_{kj}=-3\delta_{ij}$.

\section{Forward equation for the generating function of sequential functionals}
\label{app:FE}
We consider the general Langevin-equation in Ito form
\begin{equation}
\label{eq:qLangevin}
\mathrm{d}q_i = u_i \mathrm{d}t + \beta_{ij}\cdot\mathrm{d}W_j \, ,
\end{equation}
where the dot here denotes the Ito product.
The set of coordinates $q_i$ typically comprises
velocities and positions for translational motion,
but, more generally, may also contain
angular velocities and corresponding coordinates representing 
the particle orientation.
Note that we therefore adopt a slightly different notation
as in Appendix~\ref{app:EP}, where we explicitly distinguished
between space- and velocity-like coordinates.
 
The deterministic velocities are $u_i$, the $\mathrm{d}W_i$
are increments of independent Wiener processes, and the
$\beta_{ij}$ define the symmetric diffusion tensor $D_{ij}$
via $D_{ij}=\beta_{ik}\beta_{jk}$.
In the general case, $u_i$ and $\beta_{ij}$ are functions of
$q_i$ and $t$.
The Langevin-equation can also be written in Stratonovich interpretation,
\begin{equation}
\label{eq:qLangevinS}
\mathrm{d}q_i = \bar{u}_i \mathrm{d}t + \beta_{ij}\circ\mathrm{d}W_j \, ,
\end{equation}
where the relation between $u_i$ and $\bar{u}_i$ is
\begin{equation}
\label{eq:uIS}
u_i = \bar{u}_i + \frac{1}{2}\frac{\partial\beta_{ij}}{\partial q_k}\beta_{kj} \, .
\end{equation}

The general form of a functional along trajectories generated by (\ref{eq:qLangevin}) reads
\begin{equation}
\label{eq:qFunctional}
{\cal J}(\VEC{q},t|\VEC{q}_0,t_0)
= \int_{t_0}^t \left[ h\mathrm{d}t' + g_i\cdot\mathrm{d}q_i(t') + f_i\cdot\mathrm{d}W_i(t') \right] \, ,
\end{equation}
where the trajectories start at $q(0)=q_0$ at time $t_0$ and end at $q(t)=q$
at a later time $t$.
The functions $h$, $g_i$, $f_i$ depend on coordinates $q_i$ and time $t$.
The products in (\ref{eq:qFunctional}) are understood in the Ito sense.
The equivalent form of the functional with Stratonovich products is
\begin{equation}
\label{eq:qFuntionalS}
{\cal J}(\VEC{q},t|\VEC{q}_0,t_0)
= \int_{t_0}^t \left[ \bar{h}\mathrm{d}t' + \bar{g}_i\circ\mathrm{d}q_i(t') + \bar{f}_i\circ\mathrm{d}W_i(t') \right] \, .
\end{equation}
Like in (\ref{eq:qLangevinS}) we label the
the functions $\bar{h}$, $\bar{g}_i$, $\bar{f}_i$ of the Stratonovich form
by an overbar. They are
related to the Ito functions via
\begin{subequations}
\label{eq:hgfIS}
\begin{eqnarray}
h   & = & \bar{h} + \frac{1}{2}\frac{\partial\bar{g}_i}{\partial q_j} D_{ij} + \frac{1}{2}\frac{\partial\bar{f}_i}{\partial q_j}\beta_{ji}
\, , \\
g_i & = & \bar{g}_i
\, , \\[1ex]
f_i & = & \bar{f}_i
\, .
\end{eqnarray}
\end{subequations}

The generating function of ${\cal J}(\VEC{q},t|\VEC{q}_0,t_0)$ is defined as
\begin{equation}
\label{eq:qGs}
G_s(\VEC{q},t|\VEC{q}_0,t_0) = \left\langle e^{-s{\cal J}} \delta( \VEC{q}(t)-\VEC{q} ) \right\rangle \, ,
\end{equation}
with the average $\langle\ldots\rangle$ being taken over all trajectories
starting at $q_0$, $t_0$.
It can be shown
(see, for instance, Chapter 6.4 in \cite{mazo02}
or the Supplementary Material of \cite{celani12}
for the explicit derivation in special cases)
that the forward equation for the generating function $G_s$ reads
\begin{subequations}
\label{eq:qGsFK}
\begin{widetext}
\begin{equation}
\frac{\partial G_s}{\partial t} - \OP{A}^\dagger G_s =
-s \left( h + u_i g_i \right) G_s + \frac{s^2}{2} \left( D_{ij}g_i g_j + f_i f_i \right) G_s
+ s \left[ \frac{\partial}{\partial q_i} \left( D_{ij}g_j + \beta_{ij}f_j \right) G_s \right] \, ,
\end{equation}
\end{widetext}
with the usual generator of the forward diffusion process
\begin{equation}
\OP{A}^\dagger = -\frac{\partial}{\partial q_i} \left( u_i - \frac{\partial}{\partial q_j} D_{ij} \right) \, .
\end{equation}
\end{subequations}
All functions $D_{ij}$, $\beta_{ij}$, $u_i$, $h$, $g_i$, $f_i$ in (\ref{eq:qGsFK}) are evaluated at the final state $q$, $t$.

Using the definition (\ref{eq:qGs}) and integrating over this final state $q$,
we obtain
\begin{multline}
\frac{\mathrm{d}}{\mathrm{d}t} \left\langle e^{-s{\cal J}} \right\rangle
\\
= -\left\langle
\left(
s h + s u_i g_i - \frac{s^2}{2} D_{ij}g_i g_j - \frac{s^2}{2} f_i f_i
\right)
e^{-s{\cal J}}
\right\rangle \, .
\end{multline}
Finally, deriving with respect to $s$ and setting $s=0$ afterwards, we
find an equation for the average of the functional ${\cal J}$ over all trajectories
starting at $q_0$, $t_0$,
\begin{equation}
\label{eq:Jav}
\frac{\mathrm{d}}{\mathrm{d}t} \left\langle {\cal J} \right\rangle
= \left\langle h + u_i g_i \right\rangle \, .
\end{equation}

\end{document}